\documentclass[11pt]{article}

\usepackage[utf8]{inputenc}
\usepackage{amssymb,amsmath,latexsym}
\usepackage[english]{babel}
\usepackage{mathptmx}
\usepackage{graphicx}
\usepackage{subfig}
\usepackage{bibentry}
\usepackage{amsthm}
\usepackage{tabularx}
\usepackage{array}  
\usepackage[bookmarks=true,colorlinks=true,linkcolor=blue]{hyperref}
 \usepackage[colorlinks=true]{hyperref}
\hypersetup{urlcolor=blue, citecolor=red}
\usepackage{color,transparent}
 \newtheorem{thm}{Theorem}[section]
 \newtheorem{prop}{Proposition}[section]
 \newtheorem{lem}{Lemma}[section]
 
 \newtheorem{anz}{Ansatz}[section]

 \numberwithin{equation}{section}

 \def\Ai{\mathop{\rm Ai}\nolimits}
 \def\Bi{\mathop{\rm Bi}\nolimits}
\def\im{\mathop{\rm Im}\nolimits}
\def\re{\mathop{\rm Re}\nolimits}
\def\SU{\mathop{\rm SU}\nolimits}
\def\WKB{\mathop{\rm WKB}\nolimits}
\def\odd{\mathop{\rm odd}\nolimits}
\def\even{\mathop{\rm even}\nolimits}
\def\e{\mathop{\varepsilon}\nolimits}

\newcommand{\biindice}[3]%
{

\begin{array}[t]{c}
#1\\
{\scriptstyle #2}\\
{\scriptstyle #3}
\end{array}

}
\date{}
\begin{document}

\title{The flux norm, Bohr-Sommerfeld Quantization Rules and the scattering problem for $h$-$\Psi$DO's
on the real line}
\author{A.\textsc{Ifa}\,$^{1}$, M.\textsc{Rouleux}\,$^{2}$}
\maketitle
\centerline{$^{1}$ Universit\'e de Tunis El-Manar, D\'{e}partement de  Math\'{e}matiques, 1091 Tunis, Tunisia}
\centerline{\& Universit\'e de Kairouan, 3100 Kairouan, Tunisia}
\centerline{email: \url{abdelwaheb.ifa@fsm.rnu.tn}}\ \\
\centerline{$^{2}$ Universit\'{e} de Toulon, Aix Marseille Univ, CNRS, CPT, Marseille, France}
\centerline{email: \url{rouleux@univ-tln.fr}}
%\tableofcontents
%\cleardoublepage%
%\tableofcontents
%\cleardoublepage%
%\date{}%
\begin{abstract}
We revisit the well known Bohr-Sommerfeld quantization rule (BS) of order 2 for a self-adjoint 1-D $h$-Pseudo-differential
operator within the algebraic and microlocal framework of Helffer and Sj\"ostrand; BS holds precisely when
Gram matrix consisting of scalar products of some WKB solutions with respect to the ``flux norm'' (or microlocal Wonskian) is not invertible.
We simplify somewhat our previous proof \cite{IfaLouRo} by working in spatial representation only, as in complex WKB theory for
Schr\"odinger operator. We consider also the scattering problem.  
\end{abstract}
\section{Introduction}
Let $p(x,\xi;h)$ be a smooth real classical Hamiltonian on $T^*\mathbb{R}$~; we will assume that
$p$ belongs to the space of symbols $S^0(m)$ for some order function $m$ with
\begin{equation}\label{1.1}
S^{N}(m)=\big\{p \in C^{\infty}(T^{*}\mathbb{R}): \forall\,\alpha \in \mathbb{N}^{2},\,\exists\,C_{\alpha}>0,\,\forall\,(x,\xi)
\in T^{*}\mathbb{R};\; |\partial^{\alpha}_{(x,\xi)}p(x,\xi;h)|\leq C_{\alpha}\,h^{N}\,m(x,\xi)\},
\end{equation}
and has the semi-classical expansion
\begin{equation}\label{1.2}
p(x,\xi; h)\sim p_0(x,\xi)+\,hp_1(x,\xi)+\cdots,\ h\rightarrow 0.
\end{equation}
We call as usual $p_0$ the principal symbol, and $p_1$ the sub-principal symbol.
We also assume that $p+i$ is elliptic.
This allows to take Weyl quantization of $p$
\begin{equation}\label{1.3}
P(x,hD_x;h)u(x;h)=p^w(x,hD_x;h)u(x;h)=(2\,\pi\,h)^{-1}\,\int\int e^{\frac{i}{h}\,(x-y)\,\eta}\,p\big(\frac{x+y}{2},\eta;h\big)\,u(y)\,dy\,d\eta,
\end{equation}
so that $P(x,hD_x;h)$ is essentially self-adjoint on $L^2(\mathbb{R})$.
We make the geometrical hypothesis of \cite{CdV1}, namely:

Fix some compact interval $I = [E_-,E_+], E_- < E_+$, and assume
that there exists a topological ring ${\cal A}\subset p_0^{-1}(I)$ such that $\partial{\cal A} = {\cal A}_-\cup {\cal A}_+$
with ${\cal A}_\pm$ a
connected component of $p_0^{-1}(E_\pm)$.
Assume also that $p_0$ has no critical point in ${\cal A}$, and
${\cal A}_-$ is included in the disk bounded by ${\cal A}_+$ (if it is not the
case, we can always change $p$ to $-p$).

We define the microlocal well $W$ as the disk bounded by ${\cal A}_+$.
For $E\in I$, let $\gamma_E\subset W$ be a periodic orbit in the energy surface $\{p_0(x,\xi)=E\}$, so that $\gamma_E$ is an
embedded Lagrangian manifold. 

The paradigm of such an Hamiltonian, $p(x,\xi;h)=p_0(x,\xi)=\xi^2+V(x)$, is
associated with Schr\"odinger equation
\begin{equation}\label{SHRODINGER}
(P-E)u_h=\big(-h^{2}\,\Delta+V(x)-E\big)u_h=0.
\end{equation}
where $V(x)$ is a smooth ``deformation'' of $x^2$ so that ${\cal A}_\pm=\{\xi^2+V(x)=E_\pm>0\}$ are diffeomorphic to
the circles $\{\xi^2+x^2=E_\pm\}$.
We can also introduce a sub-principal symbol by considering a Sturm-Liouville equation such as $-h^2\bigl(p(x)u'(x)\bigr)'+q(x)u(x)=E$. 
Moreover $p_0(x,\xi)$ need not be invariant under $\xi\mapsto -\xi$, we can take for instance
$p_0(x,\xi)=\xi^2+f(x)\xi+V(x)$ where $f,V$ are
smooth functions. We assume that $V$ as above is even and $f$ is odd in $x$, so that $P(x,hD_x)$ verifies PT symmetry,
and the family of Lagrangian submanifolds defined by $p_0(x,\xi)=E$, look like ``shifted'' ellipses, provided
$V$ and $f$ are Morse functions, with $f(x)^2\geq4(V(x)-E)$
for all $E\in I$. This includes also higher order differential operators, with Hamiltonians like $p(x,\xi)=\xi^4+V(x)$,
used for modeling multi-layers in graphene, see e.g. \cite{GoGu}. 

As for the pseudo-differential case, we may consider Harper operator $\cos hD_x+\cos x$ on $L^2({\bf R})$ ``restricted to a potential well'',
i.e. a component of $p_0(x,\xi)=\cos\xi+\cos x=E$ where $E\in[-2,2]\setminus[-\epsilon_0,\epsilon_0]$, see \cite{HeSj}. 

We call $a=a_E=(x_E,\xi_E)\in\gamma_E$ a {\it focal point} whenever the Hamilton vector field $H_{p_0}$ turns vertical at $a$. 
If $\gamma_E$ is not a convex curve (i.e. enclosing a convex subset of $T^*{\bf R}$), it may contain many focal points. However, only the extreme ones
contribute to the quantization condition. So for simplicity we shall assume that $\gamma_E$ is convex, and thus 
contains only 2 focal points, say $a_E$ and $a'_E$, with $x'_E<x_E$.
For
$p_0(x,\xi)=\xi^2+f(x)\xi+V(x)$ such a  focal point is given by $(x_E,\xi_E)$ where $f(x_E)^2=4(V(x_E)-E)$ and $\xi_E=-f(x_E)/2$. 

If $\xi_E=0$ as in (\ref{SHRODINGER}), $x_E$ is called a {\it turning point}, and for convenience we keep this terminology in the general case. 

Then if $E_+<E_0=\displaystyle\liminf_{|x,\xi|\to\infty}p_0(x,\xi)$, all eigenvalues of $P$ in $I$
are indeed given by {\it Bohr-Sommerfeld quantization condition} (BS). 

Bohr-Sommerfeld quantization rules
hold for a general Hamiltonian (in any dimension) up to ${\cal O}(h^2)$ \cite{MaFe}, \cite{Le}, \cite{BaWe}. They are
symplectic equivariant formulas, also known as EBK or Maslov quantization rules in higher dimension.  
For general smooth 1-D Hamiltonians, BS holds
with an accuracy
${\cal O}(h^N)$ for any $N$, see \cite{Arg}, \cite{Vo77}, \cite{CdV1}, \cite{CaGra-SazLittlReiRios}.  Related results for high energy asymptotics
were obtained in \cite{HeRo82}. Their proofs are based on Functional Calculus
and Weyl correspondence (or Wigner transformation). 

In \cite{IfaLouRo} we propose microlocal asymptotics up to order $h^2$
using instead the ``flux norm'' in the spirit of \cite{Sj2} and \cite{HeSj}.
The method of the flux norm, if not the most convenient approach in the present situation,
can be suitably extended to $2\times2$ systems with branching of modes, such as Bogoliubov-de Gennes Hamiltonian \cite{PhysBogO}, \cite{PhysBog1}.
It carries to operators like Harper's operator \cite{HeSj}. As we shall see, it also applies to the scattering problem.

Exponential accuracy i.e. ${\cal O}(e^{-1/Ch})$, is obtained through the complex WKB method in the case of Schr\"odinger operator
$-h^2\Delta+V(x)$ with an analytic potential, see \cite{Dun} followed by \cite{BeOrs}, \cite{Fe}.

When computed at second order for a smooth Hamiltonian of type (\ref{1.3}) we have the following result, see e.g. \cite{CdV1} and references therein:
\begin{thm}\label{THEOREM}
Let $P(x,hD_x;h)$ be as in (\ref{1.3}). With the notations and hypotheses stated above, for $h>0$ small enough
there exists a smooth function
${\cal S}_{h}: I\to \mathbb{R}$, called the semi-classical action,
with asymptotic expansion
\begin{equation}\label{1.4}
{\cal S}_{h}(E)\sim S_0(E)+hS_1(E)+h^2S_2(E)+\cdots
\end{equation}
such that $E\in I$ is an eigenvalue of $P$ iff it satisfies the implicit equation
(Bohr-Sommerfeld quantization condition)
${\cal S}_{h}(E)=2\pi nh$, $n\in \mathbb{Z}$. The semi-classical action consists of~:

\noindent (i) the classical action along $\gamma_E$
$$S_0(E)=\oint_{\gamma_E}\xi(x)\,dx=\int\int_{\{p_0\leq E\}\cap W}\,d\xi\wedge\,dx ,$$
(ii) Maslov correction and the integral of the sub-principal 1-form $p_1\,dt$
$$S_1(E)=\pi-\int_{\gamma_E} p_1\big(x(t),\xi(t)\big)\,dt ,$$
(iii) the second order term
$$S_2(E)={1\over24}{d\over dE}\int_{\gamma_E}\Delta\, dt-
\int_{\gamma_E}p_2\,dt-{1\over2}{d\over dE}\int_{\gamma_E}p_1^2\, dt ,$$
where
$$\Delta(x,\xi)={\partial^{2} p_{0}\over\partial x^{2}}{\partial^{2} p_{0}\over\partial \xi^{2}}-
\bigl({\partial^{2} p_{0}\over\partial x\,\partial \xi}\bigr)^{2} .$$
\end{thm}
We recall that $S_3(E)=0$, as well as higher terms of odd order. Our integrals are oriented integrals, $t$ denoting the
variable in Hamilton's equations.

\noindent {\it Example}: In case of (\ref{SHRODINGER}) the semi-classical action takes the form
$${\cal S}_h(E)=\oint_{\gamma_E}\xi(x)\,dx+\pi\,h+{h^2\over12}\,{d\over dE}\int_{\gamma_E}V''\big(x(t)\big)\,dt+{\cal O}(h^4) ,$$
where $\big(\xi(x)\big)^2=E-V(x)$. The semi-classical action ${\cal S}_h(E)$ was already introduced in \cite{Dun}, see Sect.4.
Formally, it can be obtained by expressing the WKB solutions in the form
$(P_{\odd}(x;h))^{-1/2}\exp [\pm S(x,x_0;h)/h]$ (see (\ref{4.5}) below), where the phase $S(x,x_0;h)=S_0(x,x_0)+h^2S_2(x,x_0)+\cdots$ 
is such that $\partial_xS_0(x,x_0)=\xi(x)$. Then each $S_k(E)$  can be written similarly as the contour integral
$\oint_{\gamma_E}\partial_x S_k(x,x_0)\,dx$. 

A quite short proof, based on $h$-Pseudo Differential calculus, Weyl correspondence,
Moyal product or Wigner transformation,  is given in \cite{Vo77}, \cite{CdV1},
\cite{CaGra-SazLittlReiRios}, following earlier heuristic arguments by \cite{Arg}.

In \cite{IfaLouRo},  we presented instead a derivation of BS,
based on the construction of a Hermitian vector bundle of quasi-modes as in (\cite{Sj2}, \cite{HeSj}),
using different canonical charts, in the terminology of \cite{MaFe}.
Namely, if $K_h^N(E)$ denotes the microlocal kernel of $P-E$ of order $N$,
i.e. the space of microlocal solutions of $(P-E)u_h={\cal O}(h^{N+1})$ along the covering of $\gamma_E$, the problem amounts to
find the set of $E=E(h)$ such that $K_h^N(E)$ contains a global section, i.e. to construct
a sequence of quasi-modes $\big(u_n(h),E_n(h)\big)$
of a given order $N$ (practically $N=2$ or $N=4$, the third order contribution to ${\cal S}_h(E)$ vanishes).

In this paper we simplify somewhat the proof of \cite{IfaLouRo}. We also give another application of the flux norm method
to the scattering problem for Schr\"odinger operator.

The main step of \cite{IfaLouRo}
consists in computing the homology class of the semi-classical action over $\gamma_E$ up to order 2 in $h$, the leading term
being $\oint_{\gamma_E}\xi(x)\,dx$.
Our starting point was to write down the microlocal solution $\widehat u^a(\xi;h)$
near a focal point $a=(x_E,\xi_E)$ in Fourier representation mod ${\cal O}(h^2)$, see \cite{IfaLouRo} formula (3.4).
Once we know $\widehat u^a(\xi;h)$, we get the corresponding branches $u^a_\pm(x;h)$ of $u^a(x;h)$ by stationary phase
(inverse Fourier transform), in a punctured neighborhood of $a$, mod ${\cal O}(h^2)$, see \big(\cite{IfaLouRo}
formula (3.27), corrected in Erratum formula (*)\big).
We repeat the same procedure starting from the other focal point $a'=a'_E$ and then
build up Gram matrix $G^{(a,a')}(E)$ (see \cite{IfaLouRo}, formula (2.7) for a definition, and formula (\ref{2.44}) below), whose determinant,
also called {\it Jost function}, vanishes precisely
when $E$ is an eigenvalue of $P$ mod ${\cal O}(h^4)$ (there is no $h^3$ term).

Thus we have made use of 3 canonical charts starting from $a$ (one Fourier and two position representations) and another 3 starting from $a'$.
Our purpose here is to simplify the previous approach, and avoid Fourier representation,
by invoking (at least heuristically) some ideas of complex WKB method, which we make rigorous (up to order 4 in $h$, see Sect.3.2)
in case of Schr\"odinger equation (\ref{SHRODINGER}) with analytic coefficients. This reduces to 2+2 the number of canonical charts,
parametrizing the local WKB solutions $u^a_\pm(x;h)$.   

These branches meet at focal points, and 
differ by the sign of $\xi(x)-\xi_0$ in the oscillatory (or classically
allowed) region. For Schr\"odinger operator (\ref{SHRODINGER}),  the focal point is a turning point, so $\xi_0=0, V(x_0)=E$ and $\xi(x)$
take the values  $\pm\sqrt{E-V(x)}$. 

To fix the ideas, at leading order in $h$ the
microlocal solution $u$ of $(P-E)u=0$ for (\ref{SHRODINGER}) in a punctured neighborhood of $a$ takes the form (up to normalization)
\begin{equation}\label{1.5}
u^a(x,h)=\sum_\pm u^a_\pm(x;h)=e^{i\pi/4}\,(E-V)^{-1/4}\,e^{iS(a,x)/h}+e^{-i\pi/4}\,(E-V)^{-1/4}\,e^{-iS(a,x)/h}+{\cal O}(h),
\end{equation}
with the variation of Maslov index from the lower to the upper branch. So $u^a(x;h)=\sum_\pm u^a_\pm(x;h)$ is a superposition of
suitably nomalized WKB solutions, with the appropriate $e^{\pm i\pi/4}$ coefficients. 

Our claim is that this property generalizes to the $h$-PDO (\ref{1.3}).

It is justified {\it a posteriori} by formula (3.17) in \cite{IfaLouRo}, at least at second order in $h$, and
also in case of (\ref{SHRODINGER}) up to fourth order in $h$. 

Thus Fourier representation in proving Theorem \ref{THEOREM} can (formally) be avoided, but
of course it would be a difficult task to fully justify that procedure in the case of a PDO. To circumvent this difficulty, we resort to
an Ansatz, which is verified in the special case (\ref{SHRODINGER}). 

This is also related to {\it Stokes phenomenon}
for asymptotic solutions in the classically forbidden region. Consider Schr\"odinger equation (\ref{SHRODINGER})
with analytic $V$, and let $x_E$ be a simple turning point.
Let $C_E$ be Stokes curve ``outgoing'' from $x_E$, tangent
to the real axis at $x_E$. It borders (locally) Stokes regions $I\subset\{\im x<0\}$ , $II\subset\{\im x>0\}$  in the
classically forbidden region. We call $u^{a,I}_\pm(x;h)$ and $u^{a,II}_\pm(x;h)$ the corresponding asymptotics on either side of $C_E$
similar to (\ref{1.5}). 
We know that their Borel sums define two ``complex branches'' of an  exact solution
near $x_E$, which  we denote respectively by $\psi^{a,I}_\pm(x;h)$ and $\psi^{a,II}_\pm(x;h)$. 
Voros {\it connexion formula} relates the solutions $\psi^{a,I}_\pm(x;h)$ with $\psi^{a,II}_\pm(x;h)$, and is described
by the monodromy matrix $M=\begin{pmatrix}1&0\cr i&1\end{pmatrix}\in\SU(1,1)$, see \cite{Iwaki}, Theorem 1.10 and references therein. 
In constrast, on the classically allowed side, 
the solution of type (\ref{1.5}) is univalued, and the ``real branches''
$\psi^{a}_+(x;h)$ and $\psi^{a}_-(x;h)$. 
only differ by Maslov indices $e^{\pm i\pi/4}$.
It is easy to check  that
$\psi^{a}_+(x;h)$ and $\psi^{a}_-(x;h)$ are related by the matrix 
$N=\begin{pmatrix}-i&0\cr 0&i\end{pmatrix}\in \SU(2)$, (\ref{1.5}) being the asymptotics of the 
purely decaying solution in $x>x_E$ (see (\ref{3.1}) where we have switched $x_E$ and $x'_E$ to comply with the notations of \cite{Sil}.~)

Let now $P(x,hD_x;h)$ as in (\ref{1.3}) with smooth coefficents, and  $u^a_\pm(x;h)$ be the normalized asymptotic solutions
of $(P(x,hD_x;h)-E)u^a_\pm(x;h)=0$ in the classically allowed region near $a$,
ignoring the classically forbidden region. We claim that the connexion formula still holds in the asymptotic sense, namely~:

\begin{anz}\label{Anz}%Ansatz
In the classically allowed region, we have $u^a(x;h)=u^{a}_+(x;h)+u^{a}_-(x;h)$. The
  normalized asymptotic branches $u^{a}_+(x;h)$ and $u^{a}_-(x;h)$ constructed in a punctured neighborhood of the focal point $a$, to all orders in $h$,
  are related by the 
  phase factors $e^{\pm i\pi/4}$ as in (\ref{1.5}) \big(see (\ref{2.28}) below\big).
\end{anz}

In Sect.3 and 4 we check this Ansatz for Schr\"odinger operator (\ref{SHRODINGER}),
and thus recover well-known results.

Note that our constructions in \cite{IfaLouRo}, as those in  
the seminal paper \cite{Dun} based on the ``matching method'', do not use Airy function (see Sect. 3 and 4 below).
Airy functions are of course very helpful and we introduce them in Sect.3 in relation with the connexion formulas, in the sense of Voros (analytic case)
and Silvestone (smooth case). 

The paper is organized as follows~:
In Sect.\ref{Section2.1} we compute WKB solutions mod $O(h^{2})$ in the spatial representation.
They are normalized in Sect.\ref{Section2.2} using the microlocal Wronskian.
In Sect.\ref{Section2.3} we determine the homology class of the generalized action. In Sect.\ref{Section2.4} we derive Bohr-Sommerfeld quantization rule
from Ansatz (\ref{2.28}).
First two parts of this paper rely strongly on \cite{IfaLouRo}, but we have recalled the main steps of the proof for the reader's convenience,
and outlined the parallel with the present proof. 

In Sect.\ref{Section3} instead, we consider the particular case of Schr\"odinger operator with analytic coefficients.
As an alternative proof, we follow the approach of \cite{AoYo} based
on M.Sato's Microdifferential Calculus \cite{SKK}, that takes the operator to Airy operator near the turning point,
and check our claim to the fourth order in $h$.
In particular, asymptotic expansion of Airy function in a punctured neighborhood of the turning point
entails phase factors $e^{\pm i\pi/4}$, not only at leading order, but up to any accuracy in $h$. 
So $u_h^a$ is a linear combination of WKB expansions of its lower and upper branch with coefficients $e^{\pm i\pi/4}$.
We conclude by comparing our Ansatz with the ``connexion formula'', related to Stokes phenomenon, in the framework of exact WKB method.
See also \cite{Dun}, \cite{Iwaki}.

In Sect.\ref{Section4} we extend the flux norm to the problem of semi-classical scattering for Schr\"odinger operator
through a compactly supported
barrier, by computing the monodromy matrix within the framework of \cite{Ar}. For simplicity we restrict to energies above the barrier.
The case of a $h$-$\Psi$DO could be probably be handled along the same lines, see \cite{Wa} for the general approach to higher dimensions. 
%\end{document}

\medskip
\noindent {\it Acknowledgments}:
We thank Andr\'e Voros for useful remarks.

%\end{document}
\section{Quasi-modes and BS mod $\mathcal{O}(h^{2})$ in the spatial representation}%Sect.2

Here we give a proof of Theorem 1.1 along the lines of \cite{IfaLouRo} but using the Ansatz.

\subsection{WKB solutions mod $\mathcal{O}(h^{2})$ in the spatial representation}\label{Section2.1}
In a first step 
we compute smooth WKB solutions i.e. $u(x;h)=e^{i\varphi(x)/h}b(x;h)$ such that $(P-E)u(x;h)={\cal O}(h^\infty)$.
Here 
$\varphi(x)$ is the phase function verifying the eikonal equation $p_0(x,\varphi'(x))=E$, and $b(x;h)$ the amplitude, i.e. a symbol
as in (\ref{1.1}), with a formal asymptotics 
$b(x;h)=b_0(x)+hb_1(x)+\cdots$ determined recursively by solving transport equations. In the $C^\infty$ setting, $b(x;h)$ is just any Borel
sum of the $b_n(x)$.  

Actually we need several WKB solutions, we label by $u^a_\rho$, or $u^{a'}_\rho$, 
$u^{a}_{\rho}(x;h)=u^{a}_{\pm}(x;h)$ starting from the focal point $a=a_E$,
and uniformly valid
with respect to $h$ for $x$ in any $I\subset\subset]x'_E,x_E[$. Here we identify $\rho=+$ with the branch of $\gamma_E$
    connecting $a_E$ to $a'_E$ in the anti-clockwise direction, and $\rho=-$ with the other one.
Similarly $u^{a'}_\rho$ is constructed from the focal point $a'=a'_E$, we assume to be to the ``left'' of $a$. 
These WKB solutions are uniquely defined modulo their value at a given point in $]x'_E,x_E[$, or rather through a normalization procedure
(the {\it microlocal Wronskian}) (\ref{2.111}). 
    So let
\begin{equation}\label{2.1}
u^{a}_{\rho}(x;h)=b_{\rho}(x;h)\,e^{\frac{i}{h}\,\varphi_{\rho}(x)},
\end{equation}
where $b_{\rho}(x;h)$ is a formal series in $h$, which we shall compute with first order accuracy in $h$ 
\begin{equation*}
b_{\rho}(x;h)=b_{\rho,0}(x)+h\,b_{\rho,1}(x)+\cdots
\end{equation*}
%%%%%%%%%%%1
(If we think of Schr\"odinger operator, the WKB solution of (\ref{SHRODINGER}) is of the form (4.5) below, with phase function
given by $\pm S(x,x_0;h)$ where $S(x,x_0;h)$ is the generalized semi-classical action
$$S(x,x_0;h)=\int_{x_0}^x\xi(y)\,dy+h^2\int_{x_0}^x S'_2(x)\,dy+\cdots$$
So to know BS at second order in $h$, we need to compute the second correction (with accuracy $h^2$) in $(P(x,hD_x;h)-E)u(x;h)$
so the first correction $h^2S_2(x)$ in the semi-classical action, or equivalently
the first term $b_{\rho,1}(x)$ of the amplitude in (\ref{2.1}). Moreover the $S_3(x)$ term does not contribute to the homology class defining the
generalized action.

At this point we should also mention that formula (3.17) in Lemma 3.4 in \cite{IfaLouRo} has to be corrected, by removing the spurious
term $hD_2(\xi_\pm(x))$ given by (3.18). However this term is harmless since it doesn't contribute to BS.~) 

%%%%%%%%%%%1
The phase $\varphi_{\rho}(x)$ is a real smooth function that satisfies the eikonal equation
\begin{equation}\label{2.2}
p_{0}\big(x,\varphi'_{\rho}(x)\big)=E.
\end{equation}
In \cite{IfaLouRo} instead we obtain $\varphi_{\rho}(x)$ as the branches of Legendre transform $x\xi+\psi(\xi)$
near the turning points.

In case of time reversal invariance as in \label{1.4 SHRODINGER} we have $\varphi_{+}(x)=-\varphi_{-}(x)$. 
For simplicity we shall omit indices $\rho=\pm$ whenever no confusion may occur.

We look for formal solutions \big(i.e in the sense of formal classical symbols\big) of
\begin{equation}\label{2.31}
  \big(P(x,h\,D_{x};h)-E\big)\big(b(x;h)
  \,e^{\frac{i}{h}\,\varphi(x)}\big)=0 \ \Longleftrightarrow \big(Q(x,h\,D_{x};h)-E\big)b(x;h)=0
\end{equation}
where 
$Q(x,hD_x;h)=e^{-\frac{i}{h}\,\varphi(x)}P(x,hD_x)e^{\frac{i}{h}\,\varphi(x)}$ is a $h$-PDO and
\begin{equation*}
(Q-E)b(x;h)=(2\pi h)^{-1}\,\int\int e^{\frac{i}{h}\,(x-y)\,\theta}\,p\big(\frac{x+y}{2},\theta+F(x,y);h\big)\,b(y;h)\,dy\,d\theta,
\end{equation*}
with $F(x,y)=\displaystyle\int_0^1\varphi'\big(x+t(y-x)\big)\,dt$.

Applying asymptotic stationary phase at order 2, we find
\begin{equation}\label{2.3}
  \begin{aligned}
&\big(Q(x,hD_x;h)-E\big)b(x;h)=
\Big(p\big(x,\varphi'(x);h\big)-E\Big)\,b(x;h)
+\frac{h}{i}\,\Big(\beta(x;h)\,\partial_{x}b(x;h)+\frac{1}{2}\,\partial_{x}\beta(x;h)\,b(x;h)\Big)\cr
&-h^{2}\,\Big(\frac{1}{8}\,\partial_{x}r(x;h)\,b(x;h)+\frac{1}{8}\,\varphi''(x)\,\partial_{x}\theta(x;h)\,b(x;h)+\frac{1}{2}\,
\partial_{x}\gamma(x;h)\,\partial_{x}b(x;h)+\cr
&\frac{1}{2}\,\gamma(x;h)\,\frac{\partial^{2} b(x;h)}{\partial x^{2}}+\frac{1}{6}\,\varphi'''(x)\,
\theta(x;h)\,b(x;h)\Big)+\mathcal{O}(h^{3}).\cr
\end{aligned}
\end{equation}
Here $\beta(x;h), r(x;h), \theta(x;h), \gamma(x;h)$ are classical symbols,
$\beta(x;h)=\beta_0(x)+h\beta_1(x)+\cdots$, etc\dots and the first terms of their expansions are given below.

Recall $p(x,\xi;h)$ is real, $p_{0}(x_{E},\xi_{E})=E$, and $(\frac{\partial p_{0}}{\partial \xi})(x_{E},\xi_{E})\neq 0$.

Once the eikonal equation (\ref{2.2}) holds, we obtain by annihilating the term in $h$ in (\ref{2.3}) the first transport equation
\begin{equation}\label{2.4}
\beta_{0}(x)\,b'_{0}(x)+\Big(i\,p_{1}\big(x,\varphi'(x)\big)+\frac{1}{2}\,\beta_{0}'(x)\Big)\,b_{0}(x)=0,
\end{equation}
whose solutions are of the form
\begin{equation}\label{2.5}
b_{0}(x)=C_{0}\,|\beta_{0}(x)|^{-\frac{1}{2}}\,
\exp\Big(-i\,\int_{x_{E}}^{x}\frac{p_{1}\big(y,\varphi'(y)\big)}{\beta_{0}(y)}\,dy\Big),
\end{equation}
$C_0$ being so far an arbitrary constant, and $\beta_{0}(x)=(\frac{\partial p_{0}}{\partial \xi})\big(x,\varphi'(x)\big)$.
Again we have omitted the index $\rho$. 

Annihilating the term in $h^{2}$ in (\ref{2.3}), we next show that $b_{1}(x)$ is a solution of the  differential equation
\begin{equation}\label{2.6}
\begin{aligned}
&\beta_{0}(x)\,b'_{1}(x)+\Big(i\,p_{1}\big(x,\varphi'(x)\big)+\frac{1}{2}\,\beta_{0}'(x)\Big)
\,b_{1}(x)=-\beta_{1}(x)\,b'_{0}(x)-\Big(i\,p_{2}\big(x,\varphi'(x)\big)+\frac{1}{2}\,\beta'_{1}(x)\Big)\,
b_{0}(x)\cr
&+i\,\Big(\frac{1}{8}\,r'_{0}(x)\,b_{0}(x)+\frac{1}{8}\,\varphi''(x)\,\theta'_{0}(x)\,b_{0}(x)+\frac{1}{2}\,
\gamma'_{0}(x)\,b'_{0}(x)+\frac{1}{2}\,\gamma_{0}(x)\,b''_{0}(x)+\frac{1}{6}\,\varphi'''(x)\,\theta_{0}(x)\,b_{0}(x)\Big),
\end{aligned}
\end{equation}
where we have set
\begin{equation*}
r_{0}(x)=
(\frac{\partial^{3} p_{0}}{\partial x \partial \xi^{2}})\big(x,\varphi'(x)\big);\quad\gamma_{0}(x)=
(\frac{\partial^{2} p_{0}}{\partial \xi^{2}})\big(x,\varphi'(x)\big);\quad\theta_{0}(x)=(\frac{\partial^{3} p_{0}}{\partial \xi^{3}})\big(x,\varphi'(x)\big).
\end{equation*}
The homogeneous equation associated with (\ref{2.6}) is the same as (\ref{2.4});
so we are looking for a particular solution of (\ref{2.6}), integrating from $x_E$,  of the form
\begin{equation}\label{2.61}
  b_1(x)= D_{1}(x)\,|\beta_{0}(x)|^{-\frac{1}{2}}\,
  \exp\Big(-i\,\int_{x_{E}}^{x}\frac{p_{1}\big(y,\varphi'(y)\big)}{\beta_{0}(y)}\,dy\Big).
\end{equation}
Alternatively, we could integrate (\ref{2.6}) from $x'_E$ instead of $x_E$. 
So our main task will consist in computing $D_1(x)$ as a multivalued function, due to the presence of the turning points,
in the same way that we have determined $D_1(\xi)$
in \cite{IfaLouRo} (Formula (3.5)), using Fourier representation. 

We solve (\ref{2.6}) by the method of variation of constants, and find
\begin{equation}\label{2.7}
\frac{1}{C_{0}}\,\mathrm{Re}\big(D_{1}(x)\big)=-\frac{1}{2}\,\Big[\partial_{\xi}(\frac{p_{1}}{\partial_{\xi} p_{0}})\big(y,\varphi'(y)\big)\Big]_{x_{E}}^{x},
\end{equation}
\begin{equation}\label{2.8}
\begin{aligned}
&\frac{1}{C_{0}}\,\mathrm{Im}\big(D_{1}(x)\big)=\int_{x_{E}}^{x} \frac{1}{\beta_{0}}\,\bigg(-p_{2}+\frac{1}{8}\,
\frac{\partial^{4} p_{0}}{\partial y^{2} \partial \xi^{2}}+\frac{\varphi''}{12}\,\frac{\partial^{4} p_{0}}{\partial y \partial \xi^{3}}-
\frac{(\varphi'')^{2}}{24}\,\frac{\partial^{4} p_{0}}{\partial \xi^{4}}\bigg)\,dy-
\frac{1}{8}\,\int_{x_{E}}^{x} \frac{(\beta'_{0})^{2}}{\beta_{0}^{3}}\,\frac{\partial^{2} p_{0}}{\partial \xi^{2}}\,dy\cr
&+\frac{1}{6}\,\int_{x_{E}}^{x}\,\varphi''\,\frac{\beta'_{0}}{\beta_{0}^{2}}\,\frac{\partial^{3} p_{0}}{\partial \xi^{3}}\,dy+
\int_{x_{E}}^{x} \frac{p_{1}}{\beta_{0}^{2}}\,\bigg(\partial_{\xi} p_{1}-\frac{p_{1}}{2\,\beta_{0}}\,\frac{\partial^{2} p_{0}}{\partial \xi^{2}}\bigg)\,dy+
\Big[\frac{\varphi''}{6\,\beta_{0}}\,\frac{\partial^{3} p_{0}}{\partial \xi^{3}}\Big]_{x_{E}}^{x}-
\Big[\frac{\beta_{0}'}{4\,\beta_{0}^{2}}\,\frac{\partial^{2} p_{0}}{\partial \xi^{2}}\Big]_{x_{E}}^{x},
\end{aligned}
\end{equation}
Function $D_1(x)$  can be normalized by
$$D_{1}(x_{E})=0$$
These are the expressions found in \cite{IfaLouRo} (formula (3.26), corrected in the Erratum).  
The general solution of (\ref{2.6}) is given by
\begin{equation}\label{2.9}
b_{1}(x)=\big(C_{1}+D_{1}(x)\big)\,|\beta_{0}(x)|^{-\frac{1}{2}}\,
\exp\Big(-i\,\int_{x_{E}}^{x}\frac{p_{1}\big(y,\varphi'(y)\big)}{\beta_{0}(y)}\,dy\Big).
\end{equation}
It follows that
$$b(x;h)=\Big(C_{0}+h\,\big(C_{1}+D_{1}(x)\big)+\mathcal{O}(h^{2})\Big)\,|\beta_{0}(x)|^{-\frac{1}{2}}\,
\exp\Big(-i\,\int_{x_{E}}^{x}\frac{p_{1}\big(y,\varphi'(y)\big)}{\beta_{0}(y)}\,dy\Big)$$

We repeat this construction with the other branch $\rho=-1$, and thus get the 2 branches of WKB solutions
\begin{equation}\label{2.10}
u^a_{\rho}(x;h)=|\beta^{\rho}_{0}(x)|^{-\frac{1}{2}}\,
e^{\frac{i}{h}\,S_{\rho}(x_{E},x;h)}\,
\Big(C_{0}+h\,\big(C_{1}+D_{1}^{\rho}(x)\big)+\mathcal{O}(h^{2})\Big),
\end{equation}
according to (\ref{2.1}) with, omitting here the index $a$ everywhere
\begin{equation}\label{2.11}
S_{\rho}(x_{E},x;h)=\varphi_{\rho}(x_{E})+\int_{x_{E}}^{x}\xi_{\rho}(y)\,dy-h\,\int_{x_{E}}^{x}\frac{p_{1}\big(y,\varphi'_{\rho}(y)\big)}{\beta_{0}^{\rho}(y)}dy,
\end{equation}
$$\beta^{\rho}_{0}(x)=(\partial_{\xi} p_{0})\big(x,\varphi'_{\rho}(x)\big).$$
Here we have used that $\varphi_\rho(x)=\varphi_\rho(x_E)+\int_{x_E}^x\xi_\rho(y)\,dy$, where $p_0(x,\xi_\rho(x))=E$.
Note that we recover the expressions found in \cite{IfaLouRo} (formula (3.27), corrected to (*) in the Erratum).

\subsection{Well normalized quasi-modes mod $\mathcal{O}(h^{2})$ in the spatial representation}\label{Section2.2}

WKB solutions (\ref{2.10}) are not yet normalized. 
We determine the integration constants $C_{0}$, $C_{1}=C_{1}(a_{E})$ as follows.

In \cite{IfaLouRo} we called {\it microlocal Wronskian} a key invariant introduced in \cite{Sj2}, \cite{HeSj},
which allows to normalize (microlocal) WKB solutions.
Namely choose an orientation on $\gamma_E$ and denote as before by
$\rho=\pm1$ its oriented segments near $a\in\gamma_E$.
Let $\chi^a\in C_0^\infty({\bf R}^2)$ be a smooth cut-off equal to 1 near $a$, and $\omega^a_\rho$ a small neighborhood
of $\mathrm{supp} [P,\chi^a]\cap\gamma_E$ near $\rho$, where
$\chi^a$ holds for Weyl quantization $\chi^a(x,hD_x)$ as in (\ref{1.3}).
Let $u^a,v^a\in K_h(E)$ (the microlocal kernel of $P-E$) be supported microlocally on $\gamma_E$.
The microlocal Wronskian of $(u^a,\overline{v^a})$ is defined as
\begin{equation}\label{2.111}
  {\cal W}^a_\rho(u^a,\overline{v^a})=\bigl({i\over h}[P,\chi^a]_\rho u^a|v^a\bigr)
  \end{equation}
Here ${i\over h}[P,\chi^a]_\rho$ denotes the part of the
commutator supported on $\omega^a_\rho$, and $(\cdot|\cdot)$ the usual $L^2$ product.

The usual Wronskian can be seen as a ``singular limit'' of the microlocal one. 
Namely, let $P=-h^2\Delta+V$, $x_E=0$ be the turning point and change $\chi$ to Heaviside unit step-function $\chi(x)$,
depending on $x$ alone. Then in distributional sense,
we have ${i\over h}[P,\chi]=-ih\delta'+2\delta hD_x$, where $\delta$ denotes the Dirac measure at 0, and $\delta'$ its derivative,
so that $\bigl({i\over h}[P,\chi]u|u\bigr)=-ih\bigl(u'(0)\overline{u(0)}-u(0)\overline{u'(0)}\bigr)$
is the usual Wronskian of $(u,\overline u)$. However, in Sect.4 we show that the standard Wronskian is not well adapted to the
semi-classical limit.

The microlocal Wronskian of $(u^a,\overline{v^a})$ is essentially independent of the choice of the microlocal cutoff $\chi^a$, so
without loss of generality, we can
take $\chi^a(x,\xi)=\chi_1(x)\,\chi_2(\xi)$, so that $\chi_2\equiv1$ on small neighborhoods $\omega^a_\pm$, of
$\mathrm{supp}[P,\chi^{a}]\cap \{p_{0}(x,\xi)=E\}$ in $\pm(\xi-\xi_E)>0$.
Thus we need only consider the variations of $\chi_1$.

Weyl symbol of $\displaystyle \frac{i}{h}\,[P,\chi^{a}]$ is given by (we will omit index $a$ until formula (\ref{2.29}))
$$d(x,\xi;h)=d_{0}(x,\xi)+h\,d_{1}(x,\xi)+\mathcal{O}(h^{2}),$$
with
$$d_{j}(x,\xi)=\partial_{\xi}p_{j}(x,\xi)\,\chi'_{1}(x),\;\;\forall\,j\in\{0,1\}.$$
Let $$F^{a}_{\pm}=\frac{i}{h}\,[P,\chi^{a}]_{\pm}u^{a}.$$
Evaluating by stationary phase, we find
\begin{equation}\label{2.12}
\begin{aligned}
&F^{a}_{\pm}(x;h)=C_{0}\,|\beta^{\pm}_{0}(x)|^{-\frac{1}{2}}\,
\exp\bigg[\frac{i}{h}\,\Big(\varphi_{\pm}(x)-h\,\int_{x_{E}}^{x}\frac{p_{1}\big(y,\varphi'_{\pm}(y)\big)}{\beta_{0}^{\pm}(y)}dy\Big)\bigg]\cr
&\times\Big(d^{a,\pm}_{0}(x)+h\,\frac{C_{1}}{C_{0}}\,d^{a,\pm}_{0}(x)+\frac{h}{C_{0}}\,d^{a,\pm}_{0}(x)\,D^{\pm}_{1}(x)+h\,d^{a,\pm}_{1}(x)+
\frac{h}{2i}\,\big(s'_{a,\pm}(x)+2\,s_{a,\pm}(x)\,\theta_{\pm}(x)\big)+\mathcal{O}(h^{2})\Big).
\end{aligned}
\end{equation}
Here we have set
\begin{equation}\label{2.121}
  \begin{aligned}
    &d_{j}^{a,\pm}(x)=d_{j}^a\big(x,\varphi'_{\pm}(x)\big)=\beta_{j}^{\pm}(x)\,(\chi^a_{1})'(x),\;\;\forall\,j\in \{0, 1\}\cr
    &s_{a,\pm}(x)=(\partial_{\xi} d^a_{0})\big(x,\varphi'_{\pm}(x)\big)=(\frac{\partial^{2} p_{0}}{\partial \xi^{2}})\big(x,\varphi'_{\pm}(x)\big)\,(\chi^a_{1})'(x)
=\gamma^{\pm}_{0}(x)\,(\chi^a_{1})'(x)\cr
&\theta_{\pm}(x)=-\frac{1}{\beta_{0}^{\pm}(x)}\,\Big(\frac{1}{2}\,(\beta_{0}^{\pm})'(x)+i\,p_{1}\big(x,\varphi'_{\pm}(x)\big)\Big)
\end{aligned}
  \end{equation}
It follows
$$(u^{a}|F^{a}_{+})=C_{0}^{2}\,\int_{-\infty}^{x_{E}}\frac{1}{\beta_{0}^{+}}\Big(d_{0}^{a,+}+2\,h\,\frac{C_{1}}{C_{0}}\,d_{0}^{a,+}+
2\,h\,\frac{d_{0}^{a,+}}{C_{0}}\,\mathrm{Re}(D_{1}^{+})+h\,d_{1}^{a,+}+\frac{i\,h}{2}\,\big(s'_{a,+}+2\,s_{a,+}\,\overline{\theta_{+}}\,\big)\Big)\,dx.$$
It remains to show that the integral is actually independent of the cut-off $\chi_1$.

Using the fact that
$$\int_{-\infty}^{x_{E}}\frac{d_{0}^{a,+}}{\beta_{0}^{+}}\,dx=\int_{-\infty}^{x_{E}}(\chi^a_{1})'(x)\,dx=1,$$
$$\int_{-\infty}^{x_{E}}\frac{d_{1}^{a,+}}{\beta_{0}^{+}}\,dx=\int_{-\infty}^{x_{E}}\frac{\beta_{1}^{+}}{\beta_{0}^{+}}\,(\chi^a_{1})'(x)\,dx,$$
$$\int_{-\infty}^{x_{E}}\frac{d_{0}^{a,+}\,\mathrm{Re}(D_{1}^{+})}{\beta_{0}^{+}}\,dx=\frac{C_{0}}{2}\,
\partial_{\xi}(\frac{p_{1}}{\partial_{\xi} p_{0}})(a_{E})-\frac{C_{0}}{2}\,\int_{-\infty}^{x_{E}}\frac{\beta_{1}^{+}}{\beta_{0}^{+}}\,(\chi^a_{1})'(x)\,dx+
\frac{C_{0}}{2}\,\int_{-\infty}^{x_{E}}\frac{s_{a,+}\,p_{1}}{(\beta_{0}^{+})^{2}}\,dx,$$
$$\int_{-\infty}^{x_{E}}\frac{s'_{a,+}}{\beta_{0}^{+}}\,dx=\big[\frac{s_{a,+}}{\beta_{0}^{+}}\big]_{-\infty}^{x_{E}}+
\int_{-\infty}^{x_{E}}\frac{(\beta_{0}^{+})'}{(\beta_{0}^{+})^{2}}\,s_{a,+}\,dx=
\int_{-\infty}^{x_{E}}\frac{(\beta_{0}^{+})'}{(\beta_{0}^{+})^{2}}\,s_{a,+}\,dx\quad\text{(integration by parts)},$$
and
$$\int_{-\infty}^{x_{E}}\frac{s_{a,+}\,\overline{\theta_{+}}}{\beta_{0}^{+}}\,dx=
-\frac{1}{2}\,\int_{-\infty}^{x_{E}}\frac{(\beta_{0}^{+})'}{(\beta_{0}^{+})^{2}}\,s_{a,+}\,dx+
i\,\int_{-\infty}^{x_{E}}\frac{s_{a,+}\,p_{1}}{(\beta_{0}^{+})^{2}}\,dx$$

%%%%%%%%%%%3
Thus we proved the crucial
\begin{lem}%Lemma2.1
  The microlocal Wronskian 
\begin{equation}\label{2.13}
(u^{a}|F^{a}_{+})=C_{0}^{2}+h\,\Big(2\,C_{0}\,C_{1}+C_{0}^{2}\,\partial_{\xi}(\frac{p_{1}}{\partial_{\xi} p_{0}})(a_{E})\Big)+\mathcal{O}(h^{2}).
\end{equation} 
is actually independent of the cut-off $\chi_1$ mod ${\cal O}(h^2)$. 
\end{lem}
Of course we conjecture that it is independent of $\chi_1$ to every order in $h$, but this accuracy is sufficient to our purpose.
%développer un peu! C'est important car on utilise ce Lemme aussi pour le scattering.
%Ce serait bien d'insister sur la façon dont les fonction troncature disparaissent, par exemple dans le cas de
%Schrodinger, que tu as bien fait. 
%%%%%%%%%%3

Without loss of generality we may take  $C_{0}, C_{1}\in \mathbb{R}$.
Similarly
\begin{equation}\label{2.14}
(u^{a}|F^{a}_{-})=-C_{0}^{2}-h\,\Big(2\,C_{0}\,C_{1}+C_{0}^{2}\,\partial_{\xi}(\frac{p_{1}}{\partial_{\xi} p_{0}})(a_{E})\Big)+\mathcal{O}(h^{2}).
\end{equation}
\big(the mixed terms such as $(u^{a}_{\pm}|F^{a}_{\mp})$ are ${\cal O}(h^\infty)$ because the phase is non stationary\big),
so that
\begin{equation}\label{2.15}
(u^{a}|F^{a}_{+}-F^{a}_{-})=2\,C_{0}^{2}+2\,h\,\Big(2\,C_{0}\,C_{1}+C_{0}^{2}\,\partial_{\xi}(\frac{p_{1}}{\partial_{\xi} p_{0}})(a_{E})\Big)+\mathcal{O}(h^{2}).
\end{equation}

The condition that $u^{a}$ be normalized mod $\mathcal{O}(h^{2})$ is then
$${\cal W}^a_+(u^a,\overline{u^a})-{\cal W}^a_-(u^a,\overline{u^a})=(u^a|F^a_+-F^a_-)=1+{\cal O}(h^2)$$
with ${\cal W}^a_\rho(u^a,\overline{v^a})$ as in (\ref{2.111}), or

\begin{equation}\label{2.16}
C_{0}=2^{-\frac{1}{2}};\quad C_{1}=C_{1}(a_{E})=-2^{-\frac{3}{2}}\,\partial_{\xi}(\frac{p_{1}}{\partial_{\xi} p_{0}})(a_{E}).
\end{equation}
These are of course the same constants $C_0$ and $C_{1}(a_{E})$ obtained in \cite{IfaLouRo}, formula (3.6), using Fourier representation of
WKB solutions.

\subsection{The homology class of the generalized action}\label{Section2.3}%Sect.2.3

Here we identify the various terms in (\ref{2.10}), which are responsible for the holonomy of $u^a$.
Again we proceed in a very similar way to \cite{IfaLouRo} but in the position representation only.
First on $\gamma_E$ (i.e. $\Lambda_E$) we have $\varphi(x)=\displaystyle\int \xi(x)\,dx+\mathrm{Const}$. By Hamilton equations
$$\dot{\xi}(t)=-\partial_{x} p_{0}\big(x(t),\xi(t)\big),\quad
\dot{x}(t)=\partial_{\xi} p_{0}\big(x(t),\xi(t)\big),$$
so
$$\int \frac{p_{1}\big(x,\xi(x)\big)}{\beta_{0}(x)}\,dx=\int_{\gamma_{E}}\frac{p_{1}(x,\xi)}{\partial_{\xi}p_{0}(x,\xi)}\,dx=\int_{0}^{T(E)}
p_{1}\big(x(t),\xi(t)\big)\,dt.$$
The form $p_1\,dt$ is called the subprincipal 1-form. Next we consider $\sqrt{2}\,D_1(x)$ as the integral over $\gamma_E$ of the 1-form,
defined near $x_{E}$ in spatial representation, $\Omega_{1}(x)=T_{1}(x)\,dx$,
i.e. $\sqrt2D_1(x)=\int_{x_E}^xT_1(y)\,dy$.

Since $\gamma_E$ is Lagrangian, $\Omega_1$ is a closed
form that we are going to compute modulo exact forms in $\mathcal{A}$, i.e.
modulo the variations
$\Bigl[\cdot\Big]_{x_{E}}^{x}$. Using integration by parts, the integrals (\ref{2.7}), (\ref{2.8})
of $\mathrm{Re}\Omega_1(x)$, $\mathrm{Im}\Omega_1(x)$ in spatial representation simplify to
\begin{equation}\label{2.17}
\sqrt{2}\,\mathrm{Re}\big(D_{1}(x)\big)=-\frac{1}{2}\,\partial_{\xi}(\frac{p_{1}}{\partial_{\xi} p_{0}})\big(x,\xi(x)\big)-\sqrt{2}\,C_{1}(a_{E}),
\end{equation}
\begin{equation}\label{2.18}
\sqrt{2}\,\mathrm{Im}\big(D_{1}(x)\big)=\int_{x_{E}}^{x}T_{1}(y)\,dy+\Big[\frac{\varphi''}{6\,\beta_{0}}\,
\frac{\partial^{3} p_{0}}{\partial \xi^{3}}\Big]_{x_{E}}^{x}-
\Big[\frac{\beta_{0}'}{4\,\beta_{0}^{2}}\,\frac{\partial^{2} p_{0}}{\partial \xi^{2}}\Big]_{x_{E}}^{x},
\end{equation}
\begin{equation}\label{2.19}
  \begin{aligned}
T_{1}(y)&=\frac{1}{\beta_{0}}\,\bigg(-p_{2}+\frac{1}{8}\,\frac{\partial^{4} p_{0}}{\partial y^{2} \partial \xi^{2}}+
\frac{\varphi''}{12}\,\frac{\partial^{4} p_{0}}{\partial y \partial \xi^{3}}-
\frac{(\varphi'')^{2}}{24}\,\frac{\partial^{4} p_{0}}{\partial \xi^{4}}\bigg)-
\frac{1}{8}\,\frac{(\beta'_{0})^{2}}{\beta_{0}^{3}}\,\frac{\partial^{2} p_{0}}{\partial \xi^{2}}\cr
&+\frac{1}{6}\,\varphi''\,\frac{\beta'_{0}}{\beta_{0}^{2}}\,\frac{\partial^{3} p_{0}}{\partial \xi^{3}}+
\frac{p_{1}}{\beta_{0}^{2}}\,\bigg(\partial_{\xi} p_{1}-\frac{p_{1}}{2\,\beta_{0}}\,\frac{\partial^{2} p_{0}}{\partial \xi^{2}}\bigg).
\end{aligned}
\end{equation}

There follows, as in Lemma 3.2 of \cite{IfaLouRo}:
\begin{lem}%Lemma2.2
Modulo the integral of an exact form in $\mathcal{A}$, with $T_1$ as in (\ref{2.19}) we have:
\begin{equation}\label{2.20}
\begin{aligned}
&\mathrm{Re}\big(D_{1}(x)\big)\equiv 0,\cr
&\sqrt{2}\,\mathrm{Im}\big(D_{1}(x)\big)\equiv \int_{x_{E}}^{x}T_{1}(y)\,dy.
\end{aligned}
\end{equation}
\end{lem}

If $f(x,\xi), g(x,\xi)$ are smooth functions on ${\cal A}$ we set
$\Omega(x,\xi)=f(x,\xi)\, dx+g(x,\xi)\, d\xi$.
By Stokes formula
$$\int_{\gamma_{E}}\Omega(x,\xi)=\int\int_{\{p_{0}\leq E\}}\big(\partial_{x} g-\partial_{\xi} f\big)\,dx \wedge{d\xi},$$
where, following \cite{CdV1}, we have extended $p_0$ in the disk bounded by ${\cal A}_-$
so that it coincides with a harmonic oscillator in a neighborhood of a point inside, say $p_0(0,0)=0$.
Making the symplectic change of coordinates $(x,\xi)\mapsto(t,E)$ in $T^*\mathbb{R}$:
$$\int\int_{\{p_{0}\leq E\}}\big(\partial_{x} g-\partial_{\xi} f\big)\,dx \wedge{d\xi}=
\int_{0}^{E}\int_{0}^{T(E')}\big(\partial_{x} g-\partial_{\xi} f\big)\,dt \wedge{dE'},$$
where $T(E')$ is the period of the flow of Hamilton vector field $H_{p_0}$ at energy $E'$.
Taking derivative with respect to $E$, we find
\begin{equation}\label{2.21}
\frac{d}{dE}\,\int_{\gamma_{E}}\Omega(x,\xi)=\int_{0}^{T(E)}\big(\partial_{x} g-\partial_{\xi} f\big)\,dt.
\end{equation}
We compute $\displaystyle\int_{x_E}^{x} T_1(y)\,dy$ with $T_1$ as in (\ref{2.19}), and start to simplify
$J_{1}=\displaystyle \int\omega_{1}$, with $\omega_1$ the last term on the RHS of (\ref{2.19}).
Let $$g_{1}(x,\xi)=\displaystyle \frac{p_{1}^{2}(x,\xi)}{\partial_{\xi} p_{0}(x,\xi)}.$$
By (\ref{2.21}) we get
\begin{align}
J_{1}&=\frac{1}{2}\,\int_{\gamma_{E}}\frac{\partial_{\xi}\,g_{1}(x,\xi)}{\partial_{\xi}\,p_{0}(x,\xi)}\,dx
     =\frac{1}{2}\,\int_{0}^{T(E)}\partial_{\xi} g_{1}\big(x(t),\xi(t)\big)\,dt \nonumber\ \\
     &=-\frac{1}{2}\,\frac{d}{dE}\int_{\gamma_{E}}g_{1}(x,\xi)\,dx
     =-\frac{1}{2}\,\frac{d}{dE}\int_{\gamma_{E}}\frac{p_{1}^{2}(x,\xi)}{\partial_{\xi} p_{0}(x,\xi)}\,dx \nonumber\ \\
     &=-\frac{1}{2}\,\frac{d}{dE}\int_{0}^{T(E)}p_{1}^{2}\big(x(t),\xi(t)\big)\,dt, \label{2.22}
\end{align}
which is the contribution of $p_1$ to the second term $S_2$ of generalized action in (\cite{CdV1},Theorem2). Here $T(E)$ is the period on $\gamma_E$.
This is precisely the expression \cite{IfaLouRo} (Formula (3.15)) using Fourier representation. 
We also have
\begin{equation}\label{2.23}
\int_{x_{E}}^{x} \frac{1}{\beta_{0}(y)}\,p_{2}\big(y,\xi(y)\big)\,dy=
\int_{\gamma_{E}}\frac{p_{2}(x,\xi)}{\partial_{\xi}p_{0}(x,\xi)}\,dx=\int_{0}^{T(E)}p_{2}\big(x(t),\xi(t)\big)\,dt.
\end{equation}
To compute $T_1$ modulo exact forms we are left to simplify in (\ref{2.19}) the expression
\begin{align*}
J_{2}&=\int_{x_{E}}^{x} \frac{1}{\beta_{0}}\,\bigg(\frac{1}{8}\,\frac{\partial^{4} p_{0}}{\partial y^{2}
\partial \xi^{2}}+\frac{\varphi''}{12}\,\frac{\partial^{4} p_{0}}{\partial y \partial \xi^{3}}-
\frac{(\varphi'')^{2}}{24}\,\frac{\partial^{4} p_{0}}{\partial \xi^{4}}\bigg)\,dy-
\frac{1}{8}\,\int_{x_{E}}^{x} \frac{(\beta'_{0})^{2}}{\beta_{0}^{3}}\,\frac{\partial^{2} p_{0}}{\partial \xi^{2}}\,dy+\\
&\frac{1}{6}\,\int_{x_{E}}^{x}\,\varphi''\,\frac{\beta'_{0}}{\beta_{0}^{2}}\,\frac{\partial^{3} p_{0}}{\partial \xi^{3}}\,dy+
\Big[\frac{\varphi''}{6\,\beta_{0}}\,\frac{\partial^{3} p_{0}}{\partial \xi^{3}}\Big]_{x_{E}}^{x}-
\Big[\frac{\beta_{0}'}{4\,\beta_{0}^{2}}\,\frac{\partial^{2} p_{0}}{\partial \xi^{2}}\Big]_{x_{E}}^{x}.
\end{align*}
Let $g_{0}(x,\xi)=\displaystyle\frac{\Delta(x,\xi)}{\partial_{\xi} p_{0}(x,\xi)}$, where we have set according to \cite{CdV1}
$$\Delta(x,\xi)=\displaystyle\frac{\partial^{2} p_{0}}{\partial x^{2}}\,
\frac{\partial^{2} p_{0}}{\partial \xi^{2}}-\big(\frac{\partial^{2} p_{0}}{\partial x\,\partial \xi}\big)^{2}.$$
Taking second derivative of eikonal equation $p_0\big(x,\xi(x)\big)=E$, we get
\begin{equation*}
\frac{(\partial_{\xi} g_{0})\big(x,\xi(x)\big)}{\beta_{0}(x)}=
-\frac{\varphi'''}{\beta_{0}}\,\frac{\partial^{3} p_{0}}{\partial \xi^{3}}-
2\,\varphi''\,\frac{\beta'_{0}}{\beta_{0}^{2}}\,\frac{\partial^{3} p_{0}}{\partial \xi^{3}}+
\frac{\beta''_{0}}{\beta_{0}^{2}}\,\frac{\partial^{2} p_{0}}{\partial \xi^{2}}-
2\,\frac{\beta'_{0}}{\beta_{0}^{2}}\,\frac{\partial^{3} p_{0}}{\partial x\,\partial \xi^{2}}
+\frac{(\beta'_{0})^{2}}{\beta_{0}^{3}}\,\frac{\partial^{2} p_{0}}{\partial \xi^{2}}.
\end{equation*}
Integration by parts of the first and third term on the RHS gives altogether
\begin{align*}
\int_{x_{E}}^{x}\frac{(\partial_{\xi} g_{0})\big(y,\xi(y)\big)}{\beta_{0}(y)}\,dy&=
-3\,\int_{x_{E}}^{x}\frac{1}{\beta_{0}}\,\frac{\partial^{4} p_{0}}{\partial y^{2}\,\partial \xi^{2}}\,dy-
2\,\int_{x_{E}}^{x}\frac{\varphi''}{\beta_{0}}\,\frac{\partial^{4} p_{0}}{\partial y\,\partial \xi^{3}}\,dy+
\int_{x_{E}}^{x}\frac{(\varphi'')^{2}}{\beta_{0}}\,\frac{\partial^{4} p_{0}}{\partial \xi^{4}}\,dy\ \\
&+3\,\int_{x_{E}}^{x}\frac{(\beta'_{0})^{2}}{\beta_{0}^{3}}\,\frac{\partial^{2} p_{0}}{\partial \xi^{2}}\,dy
-4\,\int_{x_{E}}^{x}\varphi''\frac{\beta'_{0}}{\beta_{0}^{2}}\,\frac{\partial^{3} p_{0}}{\partial \xi^{3}}\,dy-
\big[\frac{\varphi''}{\beta_{0}}\,\frac{\partial^{3} p_{0}}{\partial \xi^{3}}\big]_{x_{E}}^{x}\ \\
&+\big[\frac{\beta'_{0}}{\beta_{0}^{2}}\,\frac{\partial^{2} p_{0}}{\partial \xi^{2}}\big]_{x_{E}}^{x}
+3\,\big[\frac{1}{\beta_{0}}\,\frac{\partial^{3} p_{0}}{\partial y\,\partial \xi^{2}}\big]_{x_{E}}^{x},
\end{align*}
and modulo the integral of an exact form in ${\cal A}$
\begin{align*}
J_{2}&\equiv-\frac{1}{24}\,\int_{x_{E}}^{x}\frac{(\partial_{\xi} g_{0})\big(y,\xi(y)\big)}{\beta_{0}(y)}\,dy
     =-\frac{1}{24}\,\int_{\gamma_{E}}\frac{\partial_{\xi} g_{0}(x,\xi)}{\partial_{\xi} p_{0}(x,\xi)}\,dx\ \\
     &=-\frac{1}{24}\,\int_{0}^{T(E)}\partial_{\xi} g_{0}\big(x(t),\xi(t)\big)\,dt
     =\frac{1}{24}\,\frac{d}{dE}\,\int_{\gamma_{E}}g_{0}(x,\xi)\,dx\ \\
     &=\frac{1}{24}\,\frac{d}{dE}\,\int_{\gamma_{E}}\frac{\Delta(x,\xi)}{\partial_{\xi} p_{0}(x,\xi)}\,dx
     =\frac{1}{24}\,\frac{d}{dE}\,\int_{0}^{T(E)}\Delta\big(x(t),\xi(t)\big)\,dt.
\end{align*}
This is again the $J_2$ computed in \cite{IfaLouRo} before Proposition 3.3, using Fourier representation. 
Using these expressions, we recover the well known action integrals (see e.g. \cite{CdV1}):
\begin{prop}\label{PROP}
Let $\Gamma\,dt$ be the restriction to $\gamma_E$ of the 1-form
$$\omega_{0}(x,\xi)=\big(\frac{\partial^{2} p_{0}}{\partial x^{2}}\,
\frac{\partial p_{0}}{\partial \xi}-\frac{\partial^{2} p_{0}}{\partial x\,\partial \xi}\,
\frac{\partial p_{0}}{\partial\,x}\big)\,dx+\big(\frac{\partial^{2} p_{0}}{\partial x\,\partial \xi}\,
\frac{\partial p_{0}}{\partial \xi}-
\frac{\partial^{2} p_{0}}{\partial \xi^{2}}\,
\frac{\partial\,p_{0}}{\partial x}\big)\,d\xi.$$
We have $\displaystyle\mathrm{Re}\oint_{\gamma_{E}}\Omega_{1}=0$, whereas
\begin{align}
\mathrm{Im}\oint_{\gamma_{E}}\Omega_{1}&=
\frac{1}{24}\,\frac{d}{dE}\,\int_{\gamma_{E}}\Delta\,dt-
\int_{\gamma_{E}}p_{2}\,dt-\frac{1}{2}\,\frac{d}{dE}
\int_{\gamma_{E}}p_{1}^{2}\,dt \label{2.24}\ \\
&=\frac{1}{48}\,\big(\frac{d}{dE}\big)^{2}\,\int_{\gamma_{E}}\Gamma\,dt-\int_{\gamma_{E}}p_{2}\,dt-
\frac{1}{2}\,\frac{d}{dE}
\int_{\gamma_{E}}p_{1}^{2}\,dt.\nonumber
\end{align}
\end{prop}

\subsection{Bohr-Sommerfeld quantization rule}\label{Section2.4}%Sect2.4
We have shown that the (normalized) WKB solutions of the eigenvalues equation $\big(P(x,hD_{x};h)-E\big)u(x;h)=0$ are given by
$u^a(x;h)=\displaystyle\sum_{\pm}u_{\pm}^a(x;h)$ as in (\ref{1.5}), where
\begin{equation}\label{2.25}
u^a_{\pm}(x;h)=|\beta_{0}^{\pm}(x)|^{-1/2}\,e^{\frac{i}{h}\,S_{\pm}(x_{E},x;h)}\,\Big(C_{0}+h\,\big(C_{1}(a_E)+D_{1}^{a,\pm}(x)\big)+\mathcal{O}(h^{2})\Big),
\end{equation}
with $C_0$ and $C_1(a_E)$ 
\begin{equation}\label{2.26}
C_{0}=2^{-1/2};\quad C_{1}(a_E)=-2^{-3/2}\,\partial_{\xi}(\frac{p_{1}}{\partial_{\xi} p_{0}})(a_{E}).
\end{equation}
the phase function $S_{\pm}(x_{E},x;h)$ is given by (\ref{2.11}), and $D_{1}^{a,\pm}(x)$ as in Lemma 2.1. 

Starting from focal point $a'_E$ instead, we can construct in a completely similar way 
\begin{equation}\label{2.27}
S_{\pm}(x'_{E},x;h)=\varphi_{\pm}(x'_{E})+\int_{x'_{E}}^{x}\xi_{\pm}(y)\,dy-h\,\int_{x'_{E}}^{x}\frac{p_{1}\big(y,\xi_{\pm}(y)\big)}{\beta_{0}^{\pm}(y)}\,dy.
\end{equation}
and the correponding symbols. So we denote again by 
$u^{a'}(x;h)=\displaystyle\sum_{\pm}u_{\pm}^{a'}(x;h)$
the microlocal solution of $\big(P(x,hD_{x};h)-E\big)u(x;h)=0$
valid uniformly
with respect to $h$ for $x$ in any $I\subset\subset]x'_E,x_E[$.

The branches labelled by $\pm$ are linearly related by some (constant) phase
factors as in the special case of Schr\"odinger operator (\ref{1.4 SHRODINGER}), see (\ref{1.5}) for the leading order term. Computing the
microlocal solutions near $a_E$ and $a'_E$ in Fourier representation as we did in \cite{IfaLouRo} shows that these phase factors are
indeed $e^{\pm i\pi/4}$. Following our Ansatz, we avoid instead this computation in inserting Maslov index $e^{\pm i \frac{\pi}{4}}$ in $u^{a}_{\pm}(x;h)$~::
\begin{equation*}
\varphi_{+}(x_{E})=\varphi_{-}(x_{E}), \  \varphi_{+}(x'_{E})=\varphi_{-}(x'_{E}),
\end{equation*}
\begin{equation}\label{2.28}
  \begin{aligned}
    &u^{a}_{\pm}(x;h)=e^{\pm i \frac{\pi}{4}}\,|\beta_{0}^{\pm}(x)|^{-1/2}\,e^{\frac{i}{h}\,S_{\pm}(x_{E},x;h)}\,
    \big(C_{0}+h\,C_{1}^{a}+h\,D_{1}^{a,\pm}(x)+\mathcal{O}(h^{2})\big),\\
    &u^{a'}_{\pm}(x;h)=e^{\mp i \frac{\pi}{4}}\,|\beta_{0}^{\pm}(x)|^{-1/2}\,e^{\frac{i}{h}\,S_{\pm}(x'_{E},x;h)}\,
    \big(C_{0}+h\,C_{1}^{a'}+h\,D_{1}^{a',\pm}(x)+\mathcal{O}(h^{2})\big).
  \end{aligned}
\end{equation}
We will justify this Ansatz in Sect.\ref{Section3} in the special case of
Schr\"odinger operator with analytic coefficients using the normal form of \cite{AoYo}. The point is that
$e^{\pm i \frac{\pi}{4}}$ are not only in factor of the principal symbol of $u^{a}_{\pm}(x;h)$, $u^{a'}_{\pm}(x;h)$, but also of the lower order terms. 

Remember (here we restore index $a$) that
\begin{equation}\label{2.29}
\begin{aligned}
&F^{a}_{\pm}(x;h)=C_{0}\,e^{\pm i \frac{\pi}{4}}\,|\beta^{\pm}_{0}(x)|^{-\frac{1}{2}}\,e^{\frac{i}{h}\,S_{\pm}(x_{E},x;h)}\cr
&\times\Big(d^{a,\pm}_{0}(x)+h\,\frac{C_{1}^{a}}{C_{0}}\,d^{a,\pm}_{0}(x)+\frac{h}{C_{0}}\,d^{a,\pm}_{0}(x)\,D^{a,\pm}_{1}(x)+h\,d^{a,\pm}_{1}(x)+
\frac{h}{2i}\,\big(s'_{a,\pm}(x)+2\,s_{a,\pm}(x)\,\theta_{a,\pm}(x)\big)+\mathcal{O}(h^{2})\Big),
\end{aligned}
\end{equation}
Similarly
\begin{equation}\label{2.30}
\begin{aligned}
&F^{a'}_{\pm}(x;h)=C_{0}\,e^{\mp i \frac{\pi}{4}}\,|\beta^{\pm}_{0}(x)|^{-\frac{1}{2}}\,e^{\frac{i}{h}\,S_{\pm}(x'_{E},x;h)}\cr
&\times\Big(d^{a',\pm}_{0}(x)+h\,\frac{C_{1}^{a'}}{C_{0}}\,d^{a',\pm}_{0}(x)+\frac{h}{C_{0}}\,d^{a',\pm}_{0}(x)\,D^{a',\pm}_{1}(x)+h\,d^{a',\pm}_{1}(x)+
\frac{h}{2i}\,\big(s'_{a',\pm}(x)+2\,s_{a',\pm}(x)\,\theta_{a',\pm}(x)\big)+\mathcal{O}(h^{2})\Big),
\end{aligned}
\end{equation}
where we recall $\theta_\pm(x)$ from (\ref{2.121}) and we have set
$$d^{a,\pm}_{0}(x)=\beta^{\pm}_{0}(x)\,(\chi_{1}^{a})'(x);\qquad d^{a',\pm}_{0}(x)=\beta^{\pm}_{0}(x)\,(\chi_{1}^{a'})'(x);\qquad \pm\beta^{\pm}_{0}(x)>0.$$
A short computation shows that modulo $\mathcal{O}(h^{2})$
\begin{equation}\label{2.31}
\begin{aligned}
(u^{a}&|F^{a'}_{+})=i\,e^{\frac{i}{h}\widetilde{A}_{+}(x_{E},x'_{E};h)}
\Big(-C_{0}^{2}-h\,C_{0}\,(C_{1}^{a}+C_{1}^{a'})+
h\,C_{0}\,\int_{x'_{E}}^{+\infty}\big(D_{1}^{a,+}+\overline{D_{1}^{a',+}}\,\big)\,(\chi_{1}^{a'})'\,dx+
h\,C_{0}^{2}\int_{x'_{E}}^{+\infty}\frac{\beta_{1}^{+}}{\beta_{0}^{+}}(\chi_{1}^{a'})'\,dx+\\
&\frac{i\,h\,C_{0}^{2}}{2}\,\int_{x'_{E}}^{+\infty}\frac{1}{\beta_{0}^{+}}\,\big(s'_{a',+}(x)+2\,s_{a',+}(x)\,\overline{\theta_{+}(x)}\,\big)\,dx\Big),
\end{aligned}
\end{equation}
where
\begin{align}
\widetilde{A}_{+}(x_{E},x'_{E};h)&=S_{+}(x_{E},x;h)-S_{+}(x'_{E},x;h)\nonumber\ \\
                                 &=\varphi_{+}(x_{E})-\varphi_{+}(x'_{E})
                                 +\int_{x_{E}}^{x'_{E}}\xi_{+}(y)\,dy-h\,\int_{x_{E}}^{x'_{E}}\frac{p_{1}\big(y,\xi_{+}(y)\big)}{\beta_{0}^{+}(y)}\,dy.
\label{2.32}
\end{align}
We know that
$$D_{1}^{a,+}(x)=-\frac{C_{0}}{2}\,\Big[\partial_{\xi}(\frac{p_{1}}{\partial_{\xi} p_{0}})\big(y,\xi_{+}(y)\big)\Big]_{x_{E}}^{x}+
i\,C_{0}\,\int_{x_{E}}^{x}T_{1}^{+}(y)\,dy+\frac{i\,C_{0}}{6}\,
\Big[\frac{\varphi''_{+}}{\beta^{+}_{0}}\,\frac{\partial^{3} p_{0}}{\partial \xi^{3}}\Big]_{x_{E}}^{x}-\frac{i\,C_{0}}{4}\,
\Big[\frac{(\beta^{+}_{0})'}{{(\beta^{+}_{0})}^{2}}\,\frac{\partial^{2} p_{0}}{\partial \xi^{2}}\Big]_{x_{E}}^{x},$$
$$D_{1}^{a',+}(x)=-\frac{C_{0}}{2}\,\Big[\partial_{\xi}(\frac{p_{1}}{\partial_{\xi} p_{0}})\big(y,\xi_{+}(y)\big)\Big]_{x'_{E}}^{x}+i\,C_{0}\,
\int_{x'_{E}}^{x}T_{1}^{+}(y)\,dy+\frac{i\,C_{0}}{6}\,
\Big[\frac{\varphi''_{+}}{\beta^{+}_{0}}\,\frac{\partial^{3} p_{0}}{\partial \xi^{3}}\Big]_{x'_{E}}^{x}-\frac{i\,C_{0}}{4}\,
\Big[\frac{(\beta^{+}_{0})'}{{(\beta^{+}_{0})}^{2}}\,\frac{\partial^{2} p_{0}}{\partial \xi^{2}}\Big]_{x'_{E}}^{x},$$
where we have set
\begin{equation}\label{2.33}
\begin{aligned}
  &T_{1}^{+}(y)=\frac{1}{\beta_{0}^{+}}\,\bigg(-p_{2}+\frac{1}{8}\,\frac{\partial^{4} p_{0}}{\partial y^{2}\,\partial \xi^{2}}+
  \frac{\varphi''_{+}}{12}\,\frac{\partial^{4} p_{0}}{\partial y\,\partial \xi^{3}}-
\frac{(\varphi''_{+})^{2}}{24}\,\frac{\partial^{4} p_{0}}{\partial \xi^{4}}\bigg)\cr
&-\frac{1}{8}\,\frac{\big((\beta^{+}_{0})'\big)^{2}}{{(\beta_{0}^{+})}^{3}}\,\frac{\partial^{2} p_{0}}{\partial \xi^{2}}+
\frac{1}{6}\,\varphi''_{+}\,\frac{(\beta_{0}^{+})'}{{(\beta_{0}^{+})}^{2}}\,\frac{\partial^{3} p_{0}}{\partial \xi^{3}}+
\frac{p_{1}}{({\beta_{0}^{+}})^{2}}\,\bigg(\partial_{\xi} p_{1}-\frac{p_{1}}{2\,\beta_{0}^{+}}\,\frac{\partial^{2} p_{0}}{\partial \xi^{2}}\bigg).
\end{aligned}
\end{equation}
Another straightforward computation shows that
\begin{align*}
\int_{x'_{E}}^{+\infty}\big(D_{1}^{a,+}(x)+\overline{D_{1}^{a',+}(x)}\big)\,(\chi_{1}^{a'})'(x)\,dx&=-C_{0}\,
\int_{x'_{E}}^{+\infty}\big(\frac{\beta_{1}^{+}}{\beta_{0}^{+}}-\frac{p_{1}}{{(\beta_{0}^{+})}^{2}}\,
\frac{\partial^{2} p_{0}}{\partial \xi^{2}}\big)\,(\chi_{1}^{a'})'(x)\,dx+
C_{1}^{a}+C_{1}^{a'}-i\,C_{0}\,\int_{x_{E}}^{x'_{E}}T_{1}^{+}(y)\,dy\ \\
&-\frac{i\,C_{0}}{6}\,\Big[\frac{\varphi''_{+}}{\beta^{+}_{0}}\,\frac{\partial^{3} p_{0}}{\partial \xi^{3}}\Big]_{x_{E}}^{x'_{E}}+\frac{i\,C_{0}}{4}\,
\Big[\frac{(\beta^{+}_{0})'}{{(\beta^{+}_{0})}^{2}}\,\frac{\partial^{2} p_{0}}{\partial \xi^{2}}\Big]_{x_{E}}^{x'_{E}}.
\end{align*}
On the other hand, integrating by parts gives
\begin{align*}
&\int_{x'_{E}}^{+\infty}\frac{s'_{a',+}(x)}{\beta^{+}_{0}(x)}\,dx=\Big[\frac{(\chi_{1}^{a'})'}{\beta^{+}_{0}}\,
\frac{\partial^{2} p_{0}}{\partial \xi^{2}}\Big]_{x'_{E}}^{+\infty}+
\int_{x'_{E}}^{+\infty}\frac{(\beta_{0}^{+})'(x)}{{\big(\beta_{0}^{+}(x)\big)}^{2}}\,
(\frac{\partial^{2} p_{0}}{\partial \xi^{2}})\big(x,\xi_{+}(x)\big)\,(\chi_{1}^{a'})'(x)\,dx=\\
&\int_{x'_{E}}^{+\infty}\frac{(\beta_{0}^{+})'(x)}{{\big(\beta_{0}^{+}(x)\big)}^{2}}\,
(\frac{\partial^{2} p_{0}}{\partial \xi^{2}})\big(x,\xi_{+}(x)\big)\,(\chi_{1}^{a'})'(x)\,dx.
\end{align*}
We also have
$$\int_{x'_{E}}^{+\infty}\frac{s_{a',+}(x)\,\overline{\theta_{+}(x)}}{\beta_{0}^{+}(x)}\,dx=
-\frac{1}{2}\,\int_{x'_{E}}^{+\infty}\frac{(\beta_{0}^{+})'(x)}{{\big(\beta_{0}^{+}(x)\big)}^{2}}\,
(\frac{\partial^{2} p_{0}}{\partial \xi^{2}})\big(x,\xi_{+}(x)\big)\,(\chi_{1}^{a'})'(x)\,dx+
i\,\int_{x'_{E}}^{+\infty}\frac{p_{1}}{{(\beta_{0}^{+})}^{2}}\,(\frac{\partial^{2} p_{0}}{\partial \xi^{2}})\big(x,\xi_{+}(x)\big)\,(\chi_{1}^{a'})'(x)\,dx,$$
and it follows that
\begin{align*}
(u^{a}|F^{a'}_{+})&\equiv i\,e^{\frac{i}{h}\widetilde{A}_{+}(x_{E},x'_{E};h)}\,\bigg(-C_{0}^{2}-i\,h\,C_{0}^{2}\,\int_{x_{E}}^{x'_{E}}T_{1}^{+}(y)\,dy
-\frac{i\,h\,C_{0}^{2}}{6}\,\Big[\frac{\varphi''_{+}}{\beta^{+}_{0}}\,\frac{\partial^{3} p_{0}}{\partial \xi^{3}}\Big]_{x_{E}}^{x'_{E}}+\frac{i\,h\,C_{0}^{2}}{4}\,
\Big[\frac{(\beta^{+}_{0})'}{{(\beta^{+}_{0})}^{2}}\,\frac{\partial^{2} p_{0}}{\partial \xi^{2}}\Big]_{x_{E}}^{x'_{E}}\bigg)
\quad \mathrm{mod}\;\mathcal{O}(h^{2})\ \\
&\equiv-\frac{i}{2}\,e^{\frac{i}{h}\,\widetilde{A}_{+}(x_{E},x'_{E};h)}\,\bigg(1+i\,h\,\int_{x_{E}}^{x'_{E}}T_{1}^{+}(y)\,dy
+\frac{i\,h}{6}\,\Big[\frac{\varphi''_{+}}{\beta^{+}_{0}}\,\frac{\partial^{3} p_{0}}{\partial \xi^{3}}\Big]_{x_{E}}^{x'_{E}}-\frac{i\,h}{4}\,
\Big[\frac{(\beta^{+}_{0})'}{{(\beta^{+}_{0})}^{2}}\,\frac{\partial^{2} p_{0}}{\partial \xi^{2}}\Big]_{x_{E}}^{x'_{E}}\bigg)
\quad \mathrm{mod}\;\mathcal{O}(h^{2}),
\end{align*}
so
\begin{equation}\label{2.35}
(u^{a}|F^{a'}_{+})\equiv-\frac{i}{2}\,e^{\frac{i}{h}\,A_{+}(x_{E},x'_{E};h)}\qquad \mathrm{mod}\;\mathcal{O}(h^{2}),
\end{equation}
and similarly
\begin{equation}\label{2.36}
(u^{a}|F^{a'}_{-})\equiv-\frac{i}{2}\,e^{\frac{i}{h}\,A_{-}(x_{E},x'_{E};h)}\qquad \mathrm{mod}\;\mathcal{O}(h^{2}),
\end{equation}
where we have set
%%%%%%%%%%4
% Il vaut mieux definir ici $A_{\pm}(x_{E},x;h)$ plutot que $A_{\pm}(x_{E},x'_{E};h)$ puisqu'on s'interesse au calcul modulo
%les formes exactes. Puis remplacer $x$ par $x'_E$.
%%%%%%%%%4
\begin{equation}\label{2.37}
  \begin{aligned}
A_{\pm}&(x_{E},x;h)=\widetilde{A}_{\pm}(x_{E},x;h)+h^{2}\,\int_{x_{E}}^{x}T_{1}^{\pm}(y)\,dy+
\frac{h^{2}}{6}\,\Big[\frac{\varphi''_{\pm}}{\beta^{\pm}_{0}}\,\frac{\partial^{3} p_{0}}{\partial \xi^{3}}\Big]_{x_{E}}^{x}
-\frac{h^{2}}{4}\,\Big[\frac{(\beta^{\pm}_{0})'}{{(\beta^{\pm}_{0})}^{2}}\,\frac{\partial^{2} p_{0}}{\partial \xi^{2}}\Big]_{x_{E}}^{x}\nonumber\ \\
&=\varphi_{\pm}(x_{E})-\varphi_{\pm}(x)+\int_{x_{E}}^{x}\xi_{\pm}(y)\,dy-h\,\int_{x_{E}}^{x}\frac{p_{1}
\big(y,\xi_{\pm}(y)\big)}{\beta_{0}^{\pm}(y)}\,dy+
h^{2}\,\int_{x_{E}}^{x'_{E}}T_{1}^{\pm}(y)\,dy+\nonumber\ \\
&\frac{h^{2}}{6}\,\Big[\frac{\varphi''_{\pm}}{\beta^{\pm}_{0}}\,\frac{\partial^{3} p_{0}}{\partial \xi^{3}}\Big]_{x_{E}}^{x}
-\frac{h^{2}}{4}\,\Big[\frac{(\beta^{\pm}_{0})'}{{(\beta^{\pm}_{0})}^{2}}\,\frac{\partial^{2} p_{0}}{\partial \xi^{2}}\Big]_{x_{E}}^{x}
  \end{aligned}
  \end{equation}
Hence we have
\begin{equation}\label{2.38}
  (u^{a}|F^{a'}_{+}-F^{a'}_{-})\equiv\frac{i}{2}\,\big(e^{\frac{i}{h}\,A_{-}(x_{E},x'_{E};h)}-e^{\frac{i}{h}\,A_{+}(x_{E},x'_{E};h)}\big)\qquad
  \mathrm{mod}\;\mathcal{O}(h^{2}).
\end{equation}
A similar computation shows that
\begin{equation}\label{2.39}
(u^{a'}|F^{a}_{+}-F^{a}_{-})\equiv\frac{i}{2}\,\big(e^{-\frac{i}{h}\,A_{-}(x_{E},x'_{E};h)}-e^{-\frac{i}{h}\,A_{+}(x_{E},x'_{E};h)}\big)\qquad \mathrm{mod}\;\mathcal{O}(h^{2}).
\end{equation}
\big(taking again into account that the mixed terms $(u^{a}_{\pm}|F^{a'}_{\mp})$ and $(u^{a'}_{\pm}|F^{a}_{\mp})$
are $\mathcal{O}(h^{\infty})$ because the phase is non stationary\big).

We conclude as in \cite{IfaLouRo}. Namely,
microlocal solutions $u^a$ and $u^{a'}$ extend as smooth solutions on the whole interval $]x'_{E},x_{E}[$;
we denote them by $u_1$ and $u_2$. Since there are no other focal points between $a$ and $a'$,
they are expressed by the same formulae (which makes the analysis particularly simple) and satisfy mod $\mathcal{O}(h^{2})$:
\begin{equation}\label{2.40}
(u_1|F^{a}_+-F^{a}_-)\equiv 1,
\end{equation}
\begin{equation}\label{2.41}
(u_2|F^{a'}_+-F^{a'}_-)\equiv -1,
\end{equation}
\begin{equation}\label{2.42}
(u_{1}|F^{a'}_{+}-F^{a'}_{-})\equiv \frac{i}{2}\,\big(e^{\frac{i}{h}\,A_{-}(x_{E},x'_{E};h)}-e^{\frac{i}{h}\,A_{+}(x_{E},x'_{E};h)}\big),
\end{equation}
\begin{equation}\label{2.43}
(u_{2}|F^{a}_{+}-F^{a}_{-})\equiv \frac{i}{2}\,\big(e^{-\frac{i}{h}\,A_{-}(x_{E},x'_{E};h)}-e^{-\frac{i}{h}\,A_{+}(x_{E},x'_{E};h)}\big).
\end{equation}
Now we define Gram matrix
\begin{equation}\label{2.44}
G^{(a,a')}(E)=\left(
       \begin{array}{cc}
         (u_1|F^{a}_+-F^{a}_-) & (u_2|F^{a}_+-F^{a}_-)\\
         (u_1|F^{a'}_+-F^{a'}_-) & (u_2|F^{a'}_+-F^{a'}_-)\\
       \end{array}
     \right),
\end{equation}
whose determinant $-\cos^{2}\big((\displaystyle{A_{-}(x_{E},x'_{E};h)-A_{+}(x_{E},x'_{E};h))/2h}\big)$ vanishes precisely on
eigenvalues of $P$ in $I$, which allows to obtain modulo $\mathcal{O}(h^{3})$
$$A_{-}(x_{E},x'_{E};h)-A_{+}(x_{E},x'_{E};h)=\pi\,h+2\,\pi\,n\,h,\;\;n\in\mathbb{Z}.$$
If $\varphi_{+}(x_{E})=\varphi_{-}(x_{E})$ and $\varphi_{+}(x'_{E})=\varphi_{-}(x'_{E})$, modulo exact forms, we have
\begin{align*}
A_{-}&(x_{E},x'_{E};h)-A_{+}(x_{E},x'_{E};h)=\int_{x'_{E}}^{x_{E}}\big(\xi_{+}(y)-\xi_{-}(y)\big)\,dy-h\,
\int_{x'_{E}}^{x_{E}}\Big(\frac{p_{1}\big(y,\xi_{+}(y)\big)}{\beta_{0}^{+}(y)}-\frac{p_{1}\big(y,\xi_{-}(y)\big)}{\beta_{0}^{-}(y)}\Big)\,dy+\\
&h^{2}\,\int_{x'_{E}}^{x_{E}}\Big(T_{1}^{+}(y)-T_{1}^{-}(y)\Big)\,dy+\mathcal{O}(h^{3}).
\end{align*}
We have
$$\int_{x'_{E}}^{x_{E}}\big(\xi_{+}(y)-\xi_{-}(y)\big)\,dy=\oint_{\gamma_{E}}\xi(y)\,dy,$$
$$\int_{x'_{E}}^{x_{E}}\Big(\frac{p_{1}\big(y,\xi_{+}(y)\big)}{\beta_{0}^{+}(y)}-\frac{p_{1}\big(y,\xi_{-}(y)\big)}{\beta_{0}^{-}(y)}\Big)\,dy=
\int_{\gamma_{E}}p_{1}\,dt,$$
\begin{align*}
\int_{x'_{E}}^{x_{E}}\Big(T_{1}^{+}(y)-T_{1}^{-}(y)\Big)\,dy&=\oint_{\gamma_{E}}T_{1}(y)\,dy
=\mathrm{Im}\oint_{\gamma_{E}}\Omega_{_{1}}(y)\ \\
&=\frac{1}{24}\,\frac{d}{dE}\,\int_{\gamma_{E}}\Delta\,dt-
\int_{\gamma_{E}}p_{2}\,dt-\frac{1}{2}\,\frac{d}{dE}
\int_{\gamma_{E}}p_{1}^{2}\,dt.
\end{align*}
This takes the proof of Theorem \ref{THEOREM} to an end.

%\end{document}
\section{Checking the Ansatz in the case of Schr\"odinger operator with analytic coefficients}\label{Section3}

We assume $V$ to be analytic near $x_E$, such that $V(x)-E\sim x-x_E$, and $x>x_E$ is the classically forbidden region (CFR).
Reduction of $P$ to its normal form $Q$ has been achieved in the framework of exact complex WKB analysis, starting from
a somewhat heuristic level in \cite{Sil} and then formalized in \cite{AoYo} using Sato's
Microdifferential Calculus, which we follow here closely. We are particularly interested in computing the precise asymptotics
of the solutions of (\ref{1.4 SHRODINGER})
up to order 4 in $h$. They are linear combinations of formal
WKB solutions $Y_{\WKB}(x;h)$ in $x>x_E$ or $x\in[x'_E,x_E]$, the classically allowed region (CAR).
The complex WKB method consists in constructing the branches
of $Y_{\WKB}(x;h)$ in the complex plane across Stokes lines, that verify (\ref{1.4 SHRODINGER}) up to exponential accuracy.
The main contribution of \cite{Sil} (where the CAR is taken instead to be $x>x'_E$) was to correct some formulas encountered
in the previous Physics literature, by taking into account the precise jump of the phase that takes place when crossing the Stokes line $\arg(x'_E-x)\sim0$.
The resulting connexion formula reads at leading order
\begin{equation}\label{3.1}
  u(x;h)\sim\big(\frac{dS}{dx}\big)^{-1/2}\big[(\tilde b-i\tilde a)e^{i\pi/4}e^{iS/h}+(\tilde b+i\tilde a)e^{-i\pi/4}e^{-iS/h}\big], \ \arg(x-x'_E)\sim0
  \end{equation}
  \begin{equation}\label{3.2}
    u(x;h)\sim\big(-\frac{d\widetilde S}{dx}\big)^{-1/2}\big[2\tilde b e^{\widetilde S/h}+(\tilde a\pm i\tilde b)e^{-\widetilde S/h}\big],  \ \arg(x'_E-x)\sim0,
    \ \mp\im x>0
  \end{equation}
Here $v=\tilde a\Ai+\tilde b\Bi$ is the general solution of $(-h^2\Delta+y)v=0$ ($\tilde a, \tilde b$ being complex constants),
$S(x;h)=S^{(0)}(x)+{\cal O}(h^2)$ is the phase with full asymptotics constructed from Ricatti equation,
see \cite{Sil}, \cite{AoYo}, \cite{Iwaki}, $S^{(0)}(x)=\displaystyle\int_{x'_E}^x\sqrt{E-V(t)}\,dt$ being the action in CAR and
$\widetilde S(x;h)=\displaystyle\int_{x'_E}^x\sqrt{V(t)-E}\,dt+{\cal O}(h^2)$, its analytic continuation in CFR.
The physical solution (purely decaying in the CFR)
is obtained with $\tilde b=0$, which we will assume here.
It is stressed in \cite{Sil} that the coefficients of the various components of (\ref{3.1})
are independent of the order in $h$ to which $S$ and $\widetilde S$ have been calculated; stated differently, it means that
(\ref{3.1}) are not only asymptotics expansions in $h$, but rather the family of analytic functions $\psi^{a',I}_\pm(x;h)$ and $\psi^{a',II}_\pm(x;h)$
indexed by $h$.

From this we derive easily the monodromy matrices $M$ and $N$ given in Ansatz 1.1 acting on coefficients $\tilde a,\tilde b\in{\bf C}$.
Namely, using (\ref{3.1}) and the fact that $V$ is real on the real domain,
we see that Voros connection formula is given by the linear operator that maps $\psi^{a',I}_\pm(x;h)$ to
$\psi^{a',II}_\pm(x;h)$, that is, ${}^t\big(2\tilde b,\tilde a-i\tilde b)$ to ${}^t(2\tilde b,\tilde a+i\tilde b\big)$, which identifies with
$M=\begin{pmatrix}1&0\cr i&1\end{pmatrix}\in\SU(1,1)$, as stated in \cite{Iwaki}, formula (1.39). A similar situation is met
for reflection over a barrier on the real line with compact support, see \cite{Ar}, Sect.5. On the other hand,
$N$ is the linear operator on ${\bf C}^2$ that maps $e^{i\pi/4}(\tilde b-i\tilde a)$ to $e^{-i\pi/4}(\tilde b+i\tilde a)\big)$, that is
$N=\begin{pmatrix}i&0\cr 0&-i\end{pmatrix}\in \SU(2)$. 

\subsection{Reduction to Airy equation}

To check Ansatz 1.1 at the level of asymptotic expansion up to order 4 in $h$,  
we take the semi-classical Schr\"odinger operator $P(x,hD_x)=-h^2\Delta+V(x)$ at energy $E$,
near a simple turning point
to Airy operator. Contrary to the standard perturbative $h$-pseudo-differential reductions to $-h^2\Delta+y$ (Egorov theorem)
we use an exact reduction by Microdifferential Calculus. 

For the reader's convenience we adopt in this Section the notations of \cite{AoYo}, and allow sometimes for a potential depending also on $h=\eta^{-1}$
as a pre-Borel summable power series of $h=\eta^{-1}$, namely $Q(\tilde x,\eta)=Q_0(\tilde x)+\eta^{-1}Q_1(\tilde x)+\cdots$.
So we need to reduce the ODE
$\big({d^2\over d\tilde x^2}-\eta^2Q(\tilde x,\eta)\big)\tilde\varphi(\tilde x,\eta)=0$
to Airy ODE $({d^2\over dx^2}-\eta^2 x)\varphi(x,\eta)=0$.

According to the prescription of Microdifferential Calculus, we identify an analytic function $\check\phi(x,y)$ with its ``symbol'', i.e.
Borel sum $\phi(x,\eta)=\int_\gamma e^{-y\eta} \check\phi(x,y)\,dy$, where $\gamma$ is an integration contour in $\re(y\eta)>0$.

This formally amounts to
quantize $\eta$ by $\partial_y$.

So we need to reduce the microdifferential operator
$\tilde A(\tilde x,\partial_{\tilde x},\partial_y)={\partial^2\over \partial\tilde x^2}-
Q(\tilde x,{\partial\over \partial y}){\partial^2\over \partial y^2}$ to the microdifferential operator
$B(x,\partial_{x},\partial_y)={\partial^2\over \partial x^2}-x{\partial^2\over \partial y^2 }$.

Recall the main result of \cite{AoYo}:
%theoreme
\begin{thm}\label{thm1}%%Theorem 3.1
Assume $Q_0(\tilde x)$ has a simple zero at $\tilde x=0$. Then, in a neighborhood of $\tilde x=0$,
and with a holomorphic change of coordinate $x(\tilde x)=x$ such that
\begin{equation}\label{3.5}
  x(\tilde x)\big(x'(\tilde x)\big)^2=Q_0(\tilde x),
  \end{equation}
with $x(0)=0$, we can find invertible microdifferential operators $S$ and $T$ with normal (ordered) product
\begin{equation}\label{3.6}
  \begin{aligned}
    &S=\big(g'(x)\big)^{5/2}\,\big(1+{\partial r(x,\eta)\over\partial x}\big)^{3/2}\,\exp\big(r(x,\eta)\xi\big):\\
    &T=\big(g'(x)\big)^{1/2}\,\big(1+{\partial r(x,\eta)\over\partial x}\big)^{-1/2}\,\exp\big(r(x,\eta)\xi\big):
  \end{aligned}
\end{equation}
such that
\begin{equation}\label{3.7}
  \tilde A(\tilde x,\partial_{\tilde x},\partial_y)\big|_{\widetilde x=g(x)}T=S\tilde B(x,\partial_{x},\partial_y).
\end{equation}
Here $\tilde x=g(x)$ denotes the inverse function of $x=x(\tilde x)$ near 0, and $r(x,\eta)\sim r_1(x)\eta^{-1}+r_2(x)\eta^{-2}+\cdots$
is a symbol of order $-1$.
\end{thm}

The ``normal (ordered) product'' of the symbol $a(x,y)\sim\sum_{i,j}a_{i,j}(x,y)\xi^i\eta^j$
consists in the quantization procedure
$$:\sum_{i,j}a_{i,j}(x,y)\xi^i\eta^j:=\sum_{i,j}a_{i,j}(x,y)\frac{\partial^i}{\partial x^i}\frac{\partial^j}{\partial y^i}$$

Note that $\exp(r(x,\eta)\xi)$ is simply the symbol of the ``shift'' operator
\begin{equation}\label{3.8}
  :\exp(r(x,\eta)\xi):\varphi(x)\big|_{x=x(\widetilde x)}=\varphi(x(\widetilde x)+r(x,\eta)),
\end{equation}
while $\big(g'(x)\big)^{1/2}\,\big(1+{\partial r(x,\eta)\over\partial x}\big)^{-1/2}$, not containing $\xi$, is (the symbol of) a multiplication operator.

Theorem \ref{thm1} is a consequence of (\cite{AoYo}, Proposition 2.2), that we use to compute the asymptotics of $r(x,\eta)$.

  Let $\tilde r(\tilde x,\eta)=x+r(x,\eta)$. We recall from (\cite{AoYo}, Eq.(2.6)) that $\tilde r(\tilde x,\eta)$ verifies the ``master'' equation
  \begin{equation}\label{3.10}
    \big({\partial\tilde r\over\partial\tilde x}\big)^2\tilde r(x,\eta)-{1\over2}\eta^{-2}\{\tilde r,\tilde x\}=
    Q_0(\tilde x)+Q_1(\tilde x)\eta^{-1}+Q_2(\tilde x)\eta^{-2}
    +\cdots.
  \end{equation}
Here $\{\tilde r,\tilde x\}$ denotes the Schwarzian derivative
\begin{equation*}
  \{\tilde r,\tilde x\}=\displaystyle\frac{\frac{\partial^{3}\tilde{r}}{\partial \tilde{x}^{3}}}{\frac{\partial \tilde{r}}
    {\partial \tilde{x}}}-\frac{3}{2}\,
\left(
  \begin{array}{c}
    \displaystyle\frac{\frac{\partial^{2}\tilde{r}}{\partial \tilde{x}^{2}}}{\frac{\partial \tilde{r}}{\partial \tilde{x}}} \\
  \end{array}
\right)^{2}.
  \end{equation*}
  Assuming that $\tilde r$ has the following asymptotics
  \begin{equation}\label{3.11}
    x+r(x,\eta)=\tilde r(\tilde x,\eta)=x_0(\tilde x)+x_1(\tilde x)\eta^{-1}+x_2(\tilde x)\eta^{-2}+x_3(\tilde x)\eta^{-3}+
    x_4(\tilde x)\eta^{-4}+\cdots,
  \end{equation}
  we find that the $x_j(\tilde x)$ solve a hierarchy of ``transport equations''. In what follows we compute the $x_j(\tilde x)$'s. 
  Denoting the differentiation with respect to $\tilde x$ by a prime, the first one (\cite{AoYo} Eq. (2.7.0))
  is of the form (\ref{3.5}), which trivially holds for $x_0=x$. So we have
\begin{equation}\label{3.12}
x(\widetilde x)=x_0(\widetilde x)=\big({3\over2}\int_0^{\tilde x}\sqrt{Q_0(\tilde y)}\,d\tilde y\big)^{2/3}.
\end{equation}
The second one (\cite{AoYo} Eq. (2.7.1)) yields the first order ODE
$$x'_0(\tilde x)\,\big(2x_0(\tilde x) {d\over d\tilde x}+x'_0(\tilde x)\big)\,x_1(\tilde x)=Q_1(\tilde x),$$
which we solve as
\begin{equation}\label{3.13}
  x_1(\widetilde x)={1\over2}\big(x_0(\widetilde x)\big)^{-1/2}\,\int^{\tilde x}\big(Q_0(\tilde y)\big)^{-1/2}\,Q_1(\tilde y)\,d\tilde y,
\end{equation}
and this vanishes when $Q_1=0$. This is consistent with the properties of WKB solution computed through Ricatti Equation.

The next ``transport equation'' (\cite{AoYo} Eq. (2.7.2)) can be written as

\begin{equation*}
x'_{0}(x'_{0}x_{2}+2x_{0}x'_{2})+x'_{1}(x'_{1}x_{0}+2x'_{0}x_{1})-\frac{1}{2}\{x_{0}, \tilde{x}\}=Q_{2}(\tilde{x}),
\end{equation*}
and we find
\begin{equation}\label{3.14}
  x_2(\widetilde x)=\frac{1}{2} \big(x_{0}(\tilde{x})\big)^{-1/2} \int^{\tilde{x}}
  \big(Q_0(\tilde y)\big)^{-1/2}\,\big(\frac{1}{2}\{x_{0},\tilde{y}\}+Q_{2}(\tilde{y})-x'_{1}(x'_{1} x_{0}+2 x'_{0}x_{1})\big)\,d\tilde{y}.
\end{equation}
In case $Q_2=0$ let us compute Taylor expansion of $x_2(\widetilde x)$ at $x=0$. We find:
\begin{equation}\label{3.15}
  x_2(\tilde x)=\frac{3}{7}v_{3}-\frac{9}{35}v_{2}^{2}+\mathcal{O}(\tilde{x}),
\end{equation}
where the coefficients $v_{j}$ are defined by:
$$Q_{0}(\tilde x)=\tilde x+\sum_{n=2}^{+\infty}v_{n}\tilde x^{n}.$$

The next ``transport equation'' (\cite{AoYo} Eq. (2.7.3)) determines $x_3(\tilde x)$. It is of the form
\begin{equation*}
  x'_{0}(x'_{0}x_{3}+2x_{0}x'_{3})+E_{1}(\tilde{x})=Q_{3}(\tilde{x}),
\end{equation*}
and we find
\begin{equation}\label{3.16}
  x_3(\widetilde x)=\frac{1}{2} \big(x_{0}(\tilde{x})\big)^{-1/2} \int^{\tilde{x}}
  \big(Q_0(\tilde y)\big)^{-1/2}\,\big(Q_{3}(\tilde{y})-E_{1}(\tilde{y})\big)\,d\tilde{y},
\end{equation}
where
$$E_{1}=2x'_{1}(x'_{0}x_{2}+x_{0}x'_{2})+x_{1}\big(2 x'_{0}x'_{2}+(x'_{1})^{2}\big)-
  \frac{1}{2}(x'_{0})^{-2}(x'_{0}x'''_{1}-x'_{1}x'''_{0})+\frac{3}{2} x''_{0}(x'_{0})^{-3}(x'_{0}x''_{1}-x'_{1}x''_{0}).$$
So again in case $Q_3(\tilde x)=0$ we have
$x_3(\tilde x)=0$.

The next ``transport equation'' (\cite{AoYo} Eq. (2.7.4)) determines $x_4(\tilde x)$, and takes the form
\begin{equation*}
x'_{0}(x'_{0}x_{4}+2x_{0}x'_{4})+E_{2}(\tilde{x})=Q_{4}(\tilde{x}),
\end{equation*}
where
$$E_{2}=2x'_{0}x'_{1}x_{3}+x_{2}\big(2x'_{0}x'_{2}+(x'_{1})^{2}\big)+
  2x_{1}\big(x'_{0}x'_{3}+x'_{1} x'_{2}\big)+x_{0}\big(2x'_{1}x'_{3}+(x'_{2})^{2}\big)-
\frac{1}{2}(x'_{0})^{-3}\,\big(x'_{0} (x'_{0} x'''_{2}-x'_{2} x'''_{0})-x'_{1}(x'_{0} x'''_{1}-x'_{1} x'''_{0})\big),$$
and we find
\begin{equation}\label{3.17}
  x_4(\widetilde x)=\frac{1}{2} \big(x_{0}(\tilde{x})\big)^{-1/2} \int^{\tilde{x}}
  \big(Q_0(\tilde y)\big)^{-1/2}\,\big(Q_{4}(\tilde{y})-E_{2}(\tilde{y})\big)\,d\tilde{y}.
\end{equation}

\subsection{WKB solution of order 4}
We use Theorem \ref{thm1} to compute $T\varphi$. Using (\ref{3.5}) and (\ref{3.6}) we see that
\begin{equation}\label{3.18}
  T\varphi(x,\eta)=\big(g'(x)\big)^{1/2}\,\big(1+{\partial r(x,\eta)\over\partial x}\big)^{-1/2}\varphi\big(x(\tilde x)+r(x,\eta)\big).
\end{equation}
If we content ourselves to the classically allowed region, we express the solution in terms of Ai function only. 
Now $\varphi(x,\eta)=\Ai(x\eta^{2/3})$ solves $\big(\frac{\partial^{2}}{\partial x^{2}}-x\frac{\partial^{2}}{\partial y^{2}}\big)
\varphi(x,\eta)=0$ (at the level of symbols). Thus by Theorem \ref{thm1} the solution of $\tilde Au=0$, evaluated at $\tilde x=g(x)$ is of the form
\begin{equation}\label{3.19}
T\varphi(x,\eta)=\big(g'(x)\big)^{1/2}\,\big(1+{\partial r(x,\eta)\over\partial x}\big)^{-1/2}\Ai \Big(\eta^{2/3}\big(x(\tilde{x})+r(x,\eta)\big)\Big).
\end{equation}
Substituting this expression in the asymptotics of Airy function in $\eta^{2/3}\,\big(x(\tilde x)+r(x,\eta)\big)=z'=\eta^{2/3}\,z<0$
gives, with $h=1/\eta$, in a punctured neighborhood of $x=0$
\begin{equation*}
  \begin{aligned}
\Ai&(z')\sim {z'}^{-1/4}\,\sin\big(\frac{2}{3h}\,{z}^{3/2}+\frac{\pi}{4}\big)\,
\big[1-{385\over4608}\,h^2\,z^{-3}+\frac{111546435}{382205952}\,h^4\,z^{-6}+{\cal O}(h^6)\big]-\\
&{z'}^{-1/4}\,\cos\big(\frac{2}{3h}\,{z}^{3/2}+\frac{\pi}{4}\big)\,
\big[{5\over48}\,h{z}^{-3/2}-{765765\over5971968}\,h^3\,{z}^{-9/2}+{\cal O}(h^5)\big]
  \end{aligned}
\end{equation*}
up to the common factor $\pi^{-1/2}$.

We expand sin and cos and factor out the phase factors $e^{\pm i\pi/4}$, so that (\ref{3.6}) gives
\begin{equation}\label{3.20}
  \begin{aligned}
  T\varphi(x,&\eta)=\frac{1}{2}\,\big(g'(x)\big)^{1/2}\,\big(x(\widetilde x)+r(x,\eta)\big)^{-1/4}\,
  \big(1+\frac{\partial r(x,\eta)}{\partial x}\big)^{-1/2}\cr
  &\Big[(-R_{2}-i\,R_{1})\,e^{i\pi/4}\,e^{\frac{2i}{3h}z^{3/2}}+(-R_{2}+i\,R_{1})\,e^{-i\pi/4}\,e^{-\frac{2i}{3h}z^{3/2}}\Big],
  \end{aligned}
\end{equation}
where
$$R_{1}(z;h)=1-\frac{385}{406}\,h^{2}\,z^{-3}+\mathcal{O}(h^{4}),$$
and
$$R_{2}(z;h)=\frac{5}{48}\,h\,z^{-3/2}+\mathcal{O}(h^{3}).$$
This shows also that Maslov correction $e^{\pm i\pi/4}$ is common to all terms of the asymptotics.
We can check also that at least, at leading order
$$\big(g'(x)\big)^{1/2}\,\big(x(\widetilde x)+r(x,\eta)\big)^{-1/4}=\big(\frac{dS}{d\tilde x}\big)^{-1/2}$$
So we can rewrite (\ref{3.20}) as (\ref{3.1}) when $\tilde b=0$, making it more precise by adding the $h^2$ correction to to
prefactors of
$e^{\pm iS(x)/h}$.
Considering the WKB solutions of (\ref{SHRODINGER}) near the focal point $a'_E$,
it is easily seen that $D_1^{a',+}(x)=\overline{D_1^{a',-}}(x)$, and $S_+(x'_E,x;h)=-S_-(x'_E,x;h)$,
so that (\ref{2.28}) agrees with (\ref{3.20}). So we have checked our Ansatz (\ref{2.28}) in case of Schr\"odinger operator,
i.e. also Ansatz 1.1, with an additional accuracy of $h^3$.

\noindent {\it Remark 1}: Asymptotics (\ref{3.11}) breaks down of course for $z$ (or $x(\tilde x)$) near 0, i.e. at the caustics.
Nevertheless the argument of Airy function is not evaluated at $x(\tilde x)$, but at $x(\tilde x)+r(x,\eta)$. Looking at the asymtotics
(\ref{3.5})  we see that if $x_2(0)<0$, then $x(\tilde x)+r(x,\eta)=x(\tilde x)+h^2x_2(\tilde x)+\cdots$ is $\sim -h^2$
already for $x(\tilde x)=0$. So asymptotics (\ref{3.11}) is indeed ``regular'' at $x(\tilde x)=0$, and exact WKB method ``smears out''
the geometric singularity given by ordinary asymptotics in that case. By
(\ref{3.8}) this holds when $v_3=0$. On the contrary, if $x_2(0)>0$, asymptotics (\ref{3.11}) is already ``singular'' for
$x(\tilde x)\sim h^2$. These effects of course are largely irrelevant from the point of vue of geometric asymptotics.

%\end{document}
%\input Michel-Abd4.tex
%\end{document}

\section{Scattering over a barrier: case of a compactly supported smooth potential}\label{Section4}

The scattering problem for Schr\"odinger equation $-h^2u''(x)+(V(x)-E)u(x)=0$ (\ref{1.4 SHRODINGER})
at a non-trapping energy $E$ has received considerable attention,
depending on the regularity of the potential and its behavior at infinity.

The simplest case of a compactly supported,
piecewise continuous potential $V$ for which standard ODE technics apply (like Wronskians) has been considered in \cite{Ar}.

Define the {\it monodromy operator} of (\ref{1.4 SHRODINGER}) 
as the linear operator $M(h)$ mapping the state space of a particle with energy $E=k^2>0$
into itself as follows.  To a solution of Schr\"odinger operator for a free particle
$-h^2u''(x)-k^2u(x)=0$ we assign a solution of (\ref{1.4 SHRODINGER}) coinciding with it to the left of the support,
and to this solution, in turn, we assign its value to the right of the support.
It is shown in \cite{Ar} that (\ref{1.4 SHRODINGER}) has a unique solution $u$ (which we call Arnold solution) such that
$u(x)=e^{ikx}+Be^{-ikx}$ to the left of the barrier,
and $u(x)=Ae^{ikx}$ to the right of the barrier.

Then the matrix of the monodromy operator in the basis $(f_d,f_g)$
such that $f_d(x)=e^{ikx}$ (wave travelling to the right) and $f_g(x)=e^{-ikx}$ (wave travelling to the left) belongs to the group $\SU(1,1)$.
It is of the form
\begin{equation}\label{4.1}
  M(E,h)=\begin{pmatrix}1/\overline A&-\overline B/\overline A\\-B/A&1/A\end{pmatrix}
  \end{equation}
where the complex numbers $A,B$ such that $|A|^2+|B|^2=1$
are called the {\it transmission}
and the {\it reflexion} coefficients. 

When $V$ is a step function (``hard wall potential'') it is easy to compute the probability amplitudes $A$
and $B$ using that the solution is $C^1$, see \cite{Ar}.
For large energies (above the maximum $V_0$ of $V$, i.e. the height of the
barrier), the reflection coefficient $|B|^2$ tends to 0, while for energies less than $V_0$, the transmission coefficient $|A|^2$ is exponentially small
(because of tunneling).  

When $V$ instead is a smooth potential tending sufficiently fast to some limits $V_-$ as $x\to -\infty$, and $V_+$ as $x\to+\infty$ 
see \cite{LaLi}, then (\ref{1.4 SHRODINGER}) has 2 families of fundamental solutions $(u_d^+, u_g^+)$ and $(u_d^-,u_g^-)$ with the following WKB asymptotics

\begin{equation}\label{4.2}
  \begin{aligned}
    u_d^+(x,h)=b^+_d(x;h)\exp [i\int_{x_1}^x\varphi'(x')\,dx'/h], \quad x\to+\infty\cr
    u_d^-(x,h)=b_d^-(x,h)\exp [i\int_{x_2}^x\varphi'(x')\,dx'/h], \quad x\to-\infty\cr
  \end{aligned}
  \end{equation}
    and
\begin{equation}\label{4.3}
  \begin{aligned}
    u_g^+(x,h)=b_g^+(x,h)\exp [-i\int_{x_2}^x\varphi'(x')\,dx'/h], \quad x\to+\infty\cr
    u_g^-(x,h)=b_g^-(x,h)\exp [-i\int_{x_1}^x\varphi'(x')\,dx'/h], \quad x\to-\infty\cr
  \end{aligned}
  \end{equation}
where $x_1<<0$ and $x_2>>0$ are some ``base points'' to be defined below. 
The scattering matrix $S(h)=\begin{pmatrix}s_{11}&s_{12}\\s_{21}&s_{22}\end{pmatrix}$ is defined as follows.
Any solution of (\ref{1.4 SHRODINGER}) takes the form
\begin{equation}\label{4.4}
  u=c_d^-u_d^-+c_g^-u_g^-=c_d^+u_d^++c_g^+u_g^+
\end{equation}
the complex coefficients $c_d^\pm$, $c_g^\pm$ being related by $\begin{pmatrix}c_d^+\\c_g^-\end{pmatrix}=S(h)\begin{pmatrix}c_d^-\\c_g^+\end{pmatrix}$,
    see \cite{Fe}.
    When $V=0$ we can take up to constant factors, $(u_d^+, u_g^+)=(u_d^-, u_g^-)=(e^{ikx/h},e^{-ikx/h})$, so that $S(h)=I$. 
    More generally, for $(c_d^-,c_g^+)=(1,0)$ (Arnold solution),
the monodromy operator $M(E,h)$ (\ref{4.1})  is simply related with the scattering matrix $S(h)$.

In case $V_\pm=0$ and a short range analytic potential, the construction of exact solutions can be carried out as in \cite{GeGr},
using analytic extensions in the complex plane, so computing 
Wronskians and their asymptotics leads to $S(E,h)$ (see \cite{Ra}). However, the Wronskian of the (exact) solutions
is not asymptotic to the Wronskian of their asymptotics.
Consider for instance WKB solution in the
form
  $\psi_\pm(x;h)=\exp [\int^x P^\pm(x';h)\,dx'/h]$
(see \cite{Iwaki}), where $P^\pm(x;h)\sim\sum_{m\geq-1} h^mP_m^\pm(x)$. Decomposing $P^\pm(x;h)$
into its odd and even parts, resp. $P_{\odd}=(P^+-P^-)/2$ and $P_{\even}=(P^++P^-)/2$, choosing some base point $x_0$, we get
\begin{equation}\label{4.5}
 \psi_\pm(x;h)=(P_{\odd}(x;h))^{-1/2}\exp [\pm\int_{x_0}^x P_{\odd}(x';h)\,dx'/h]= (P_{\odd}(x;h))^{-1/2}\exp [\pm S(x,x_0;h)/h]
\end{equation}  
which is the expression used in (\ref{3.1}), see e.g. \cite{Iwaki}. Formally the Wronskian of $\psi^\pm$ would be
$W(\psi_+,\psi_-)=\psi'_+\psi_--\psi'_-\psi_+$, and using that $P_{\even}(x;h)=-\frac{1}{2}\frac{d}{dx}\log P_{\odd}(x;h)$, we find
$W(\psi_+,\psi_-)=2\sqrt{P_{\odd}(x;h)}\neq0$ and this not a constant, except when $x$ lies outside of the support of $V$.
If we use instead the asymptotic representation
$\psi_\pm(x;h)=(a_0(x)+ha_1(x)+\cdots)\exp[\pm S(x,x_0;h)/h]$, we find $W(\psi_+,\psi_-)=2+hW_1(x)+\cdots$
where again $W_1(x)$ is not a constant. 

However, we will show that replacing the usual Wronskians by the ``microlocal Wronskians'' (\ref{2.111}) gives the correct interpretation. 

As a warm-up let us show how to retrieve the identity $|A|^2+|B|^2=1$ between the transmission and reflection  coefficients.

Let $\widetilde \chi\in C^\infty_0({\bf R})$, and $\widetilde\chi(x,hD_x)$ its Weyl quantization. For any solutions 
$u,v\in C^\infty({\bf R})$ of Schr\"odinger equation $P(x,hD_x)u=Eu$, we define as in (\ref{2.111}) their microlocal Wronskian by
${\cal W}(u,v)=(\frac{i}{h}[P,\widetilde\chi]u|v)$. Since $u,v$ are smooth and $\widetilde\chi$ compactly supported, we have
\begin{equation}\label{4.6}
  (\frac{i}{h}[P,\widetilde\chi]u|v)=0
  \end{equation}
so in particular the microlocal Wronskian is independent of $\widetilde\chi$.
Let also $a',a\in T^*{\bf R}$ project outside of the support of $V$, and belong to the characteristic set $\xi^2+V(x)-E=0$ with $\xi=\varphi'(x)=k>0$.
Let $b',b$ denote their image by the reflection to the $x$ axis, with $\xi=-\varphi'(x)$.

Thus the projection $x'_0=x(a')=x(b')$ of $a',b'$ lies to the left of the support of $V$, while
$x_0=x(a)=x(b)$ lies to the right of the support of $V$. Let $\widetilde\chi_0$ be a cut-off in ${\bf R}^2$,
such that the support of $\nabla\widetilde\chi_0$ intersects the characteristic set $\xi^2+V(x)-E=0$ at $a,a',b',b$.
So Arnold solution $u$ is microlocally equal to $Ae^{ikx/h}$, $e^{ikx/h}$, $Be^{-ikx/h}$ and 0 near $a,a',b',b$
respectively (we recall
$k=\sqrt E$). As before we denote by $\frac{i}{h}[P,\widetilde\chi]_a$ the contribution of $\frac{i}{h}[P,\widetilde\chi]$
near $a$, etc\dots Choose also $\chi_0\in C^\infty_0({\bf R})$ with $\chi_0(x'_0)=\chi_0(x_0)=1$ 
be such that $\widetilde\chi_0(x,\xi)=\chi_0(x)$ (independent of $\xi$)
near $a,a',b,b'$. By (\ref{4.6}) we have 

\begin{equation}\label{4.7}
  (\frac{i}{h}[P,\widetilde\chi_0]_{a'}u|u)+(\frac{i}{h}[P,\widetilde\chi_0]_au|u)+(\frac{i}{h}[P,\widetilde\chi_0]_bu|u)+
  (\frac{i}{h}[P,\widetilde\chi_0]_{b'}u|u)=0
  \end{equation}
Using $\frac{i}{h}[P,\widetilde\chi]=-ih\chi''_0(x)+2\chi'_0(x)hD_x$ we find, since $u=0$ microlocally near $b$
\begin{equation}\label{4.8}
  \begin{aligned}
  \int_{-\infty}^{x'_0}&\frac{h}{i}\chi''_0(x)\,dx+\int_{-\infty}^{x'_0}2\chi'_0(x)k\,dx+\int_{x_0}^\infty \frac{h}{i}\chi''_0(x)\,dx+\cr
  &\int_{x_0}^\infty 2\chi'_0(x)k|A|^2\,dx+\int_{-\infty}^{x'_0}\frac{h}{i}\chi''_0(x)|B|^2\,dx+\int_{-\infty}^{x'_0}2\chi'_0(x)(-k|B|^2)\,dx=0
  \end{aligned}
\end{equation}
and by integration, $0=2k-2k|A|^2-2k|B|^2$ which gives the result (conservation of the probability flux).
Actually, all equalities above hold mod ${\cal O}(h^\infty)$,
because of the asymptotic character of Weyl quantization, but the relation $|A|^2+|B|^2=1$ turns out to be exact.

We shall now consider energies $E>V_0$ above the barrier, so $P(x,hD_x)$ has simple characteristics and every WKB solution can be smoothly extended
to the entire real line. In other words, modes travelling to the right or to the left do not mix, and for Arnold solution, we have $B={\cal O}(h^\infty)$,
$|A|=1+{\cal O}(h^\infty)$.  
So we are left to compute $A$, which reduces to a phase factor.

The phase $\varphi$ satisfies the eikonal equation
$\varphi'(x)^2+V-E=0$ (where we can choose $\varphi'(x)>0$ for all $x\in{\bf R}$), and $b_\rho^\pm(x,h)$ are amplitudes with common leading term
$b_0(x)=(E-V(x))^{-1/4}$  (see (\ref{2.1}) and (\ref{1.5}), where we can think here of the turning points $a,a'$ being mapped to $\pm\infty$).  

By (\ref{1.4}) we have mod ${\cal O}(h^\infty)$
$$u(x;h)=c_d^+u_d^+(x;h)=u_d^-(x;h)=c_d^-u_d^-(x;h)=Au_d^+(x;h)$$
hence $c_d^-=1$, and $c_d^+=\frac{u_d^-}{u_d^+}(x;h)$. In (\ref{4.2}) the asymptotics are evaluated at $\pm\infty$, but
the real amplitudes $b_d^\pm(x;h)$ take the same form, at least up to ${\cal O}(h)$. 
We let the base points $x_1\to-\infty$, $x_2\to+\infty$ and regularize the integrals defining the phase function $\varphi(x)$.
As in \cite{Fe} (formula (9), Sect.11) we end up with
\begin{equation}\label{4.9}
  A=\exp[i\int_{\bf R}(\sqrt{E-V(x)}-\sqrt E)\,dx/h]+{\cal O}(h)
  \end{equation}

In case $0<E<\max V$, the situation is more complicated since the modes $u_d(x;h)$ and $u_g(x;h)$ do not decouple. 
In particular we have tunneling through the barrier $[E, \max V]$, where the phase becomes purely imaginary. 
But we can proceed as before in the classically allowed region. Namely, let 
$\widetilde\chi\in C_0^\infty(T^*{\bf R})$ whose support intersects $\xi^2+V(x)-E=0$,
we can define the set of localized functions $(F_1^+,F_1^-),(F_2^+,F_2^-)$ by
\begin{equation}\label{4.11}
  \begin{aligned}
    &F_2^+=\frac{i}{h}[P,\widetilde\chi]_au_d^+, \quad F_1^+=\frac{i}{h}[P,\widetilde\chi]_{b'}u_g^+\cr
    &F_1^-=\frac{i}{h}[P,\widetilde\chi]_{a'}u_d^-, \quad F_2^-=\frac{i}{h}[P,\widetilde\chi]_bu_g^-
    \end{aligned}
\end{equation}
and by analogy with Sect.2, we form the linear combinations $(F_2^++\e _2F_2^-,F_1^++\e _1F_1^-)$, with indices $\e _j=\pm1$
to be chosen lateron, which give a set of basis of the microlocal co-kernel of $P(x,hD_x)-E$ in the classically allowed region.
Changing $\xi$ to $i\xi$ allows to extend WKB solutions in the classically forbidden region.
Matching near the boundary of the barrier can be carried out as in Sect.3, and lead to the connexion formula.
Thus we the monodromy matrix $M(E,h)\in \SU(1,1)$ as a product of ``local monodromy matrices'' of the type of Sect.3.
This will be investigated elsewhere.

%\end{document}

%\end{document}

%bibliographie
 %\nocite{*}
 %\bibliographystyle{plain}
 %\bibliography{Michel-Abd}

\begin{thebibliography}{}

  \bibitem{AoYo} %Airy function
T. Aoki and J. Yoshida. Microlocal reduction of ordinary differential operators with a large parameter. Publ RIMS, Kyoto
Univ, 29:959–975, 1993.

  \bibitem{Arg}
    P.N. Argyres. The Bohr-Sommerfeld Quantization Rule and the Weyl correspondence. Physics, (2):131–199, 1965.

    \bibitem{Ar}
      V.Arnold. Geometrical methods in the theory of ODE's. Springer, 1983.

    \bibitem{AssFuHi}
      M.Assal, S.Fujiie, K.Higuchi. Transition of the semiclassical resonance widths across a tangential crossing energy-level.
      J. Math. Pures Appl. 191(3-4), 2024.
      
 \bibitem{BaWe}
  S.Bates, A.Weinstein. Lectures on the geometry of quantization. Berkeley Math. Lect. Notes 88,
  American Math. Soc. 1997.
  
  \bibitem{BeOrs}
    C. Bender and S. Orszag. Advanced Mathematical Methods for scientists and engineers. Springer, 1979.
  
 \bibitem{BoFuRaZe}
  J.-F. Bony, S.Fujiie, T.Ramond, M.Zerzeri. Quantum monodromy for a homoclinic orbit. Proc.
Colloque EDP Hammamet, 2003. 
 
\bibitem{CaGra-SazLittlReiRios}
  M.Cargo, A.Gracia-Saz, R.Littlejohn, M.Reinsch \& P.de Rios. 
  Moyal star product approach to the Bohr-Sommerfeld approximation, J.Phys.A: Math and Gen.38, p.1977-2004, 2005.


    \bibitem{CdV1}
  Y. Colin de Verdi\`ere. Bohr-Sommerfeld Rules to All orders. Ann. H.Poincar\'e, 6:925–936, 2005.

  
  \bibitem{Dun}
J. L. Dunham. The WKB Method of Solving the Wave Equation. Phys. Rev, 41(713), 1932.

\bibitem{DuRay}
A.Duraffour, N.Raymond. 
  An Example of Accurate Microlocal Tunneling in One Dimension. arXiv:2407.03747v3
 
\bibitem{Fe}
  M.V. Fedoriouk. M\'ethodes asymptotiques pour les \'{E}quations Diff\'erentielles Ordinaires Lin\'eaires. Asymptotic Analysis.
Springer, Moscou, Mir edition, 1987.

\bibitem{GeGr}
  C.G\'erard, A.Grigis. Precise estimates of tunneling and eigenvalues near a potential barrier. J. Diff. Equations 42, p.149-177, 1988.

\bibitem{GoGu}
  E.Gorbar, V.Gusynin. Bound states of quasiparticles with quartic dispersion in an external potential~: WKB
  approach. arXiv:2502.13616.

  \bibitem{HeRo82}
  B.Helffer, D.Robert. Asymptotique des niveaux d'\'energie pour des Hamiltoniens \`a un degr\'e de libert\'e.
  Duke Math. J. Vol.49 (4), p.853-868, 1982.

  \bibitem{HeSj1}
    B. Helffer and J. Sj\"{o}strand. Multiple wells in the semi-classical limit. CPDE, 1986.
    
\bibitem{HeSj}
  B. Helffer and J. Sj\"{o}strand. Semi-classical analysis for Harper's equation III. Mémoire No 39, Soc. Math. de France, 117(4), 1988.
  
\bibitem{Ifa}
A. Ifa. Bohr-Sommerfeld Quantization conditions for Schrodinger operator: the method of microlocal Wronskian and Gram matrix. arXiv:2509.23514 

\bibitem{IfaLouRo}%1
  A. Ifa H. Louati and M. Rouleux. Bohr-Sommerfeld Quantization Rules Revisited: the Method of Positive Commutators.
  J. Math. Sci. Univ. Tokyo, 25(2):1–37, 2018. Erratum: J. Math. Sci. Univ. Tokyo, 27(1):81–85, 2020.

  \bibitem{IfaM'haRo}
  A. Ifa N. M'hadbi and M. Rouleux. On generalized Bohr-Sommerfeld quantization rules for operators with PT symmetry.
  Mathematical Notes, 99(5):673–683, 2016.
  
  \bibitem{PhysBogO}
A. Ifa and M. Rouleux. The one dimensional semi-classical Bogoliubov-de Gennes Hamiltonian with PT symmetry:
generalized Bohr-Sommerfeld quantization rules. J. of Physics: Conf. Series, 1194:1–11, 2019.

\bibitem{PhysBog1}
  A. Ifa and M. Rouleux. The one-dimensional semi-classical Bogoliubov-de Gennes Hamiltonian.
hal-02072721. Submitted

\bibitem{Iwaki}
  K.Iwaki. Les Houches Lectures on
  Exact WKB Method and Painlev\'e Equations, 2024.
  

\bibitem{LaLi}
  L.Landau, E.Lifschitz. {\it Physique Th\'eorique, Vol.3, M\'ecanique Quantique}. Mir, Moscou, 1967.

\bibitem{Le}
  J.Leray. Analyse Lagrangienne et M\'ecanique Quantique. S\'eminaire EDP Coll\`ege de France, 1976-1977.

\bibitem{MaFe}
V.P. Maslov and M.V. Fedoriuk. Semi-classical approximation in Quantum Mechanics. D. Reidel Publishing Company, 1981.

\bibitem{Ol}
F. Olver. Introduction to Asymptotics and Special Functions. Academic Press, New York, 1974.

  \bibitem{Ra}
    T.Ramond. Semiclassical Study of Quantum Scattering on the Line. Comm. Math. Phys.177, p.221-254, 1996.
    
\bibitem{SKK}
M. Sato T. Kawai and M. Kashiwara. Microfunctions and pseudo-differential equations. Lecture Notes in Math., Springer,
287:265–529, 1973.

\bibitem{Sil}
H. Silverstone. JWKB Connection-Formula Problem revisited via Borel Summation. Physical Rev. Lett.,
55 (23):2523–2526, 1986.

\bibitem{Sj2}
J. Sj\"{o}strand. Density of states oscillations for magnetic Schr\"{o}dinger operators, in: Bennewitz (ed.). Diff. Eq. Math. Phys,
p.295–345, 1990.


\bibitem{Vo77}
  A. Voros. Asymptotic $h$-expansions of stationary quantum states. Ann. Inst. Henri Poincar\'e, A26, p.343-403, 1977.

\bibitem{Wa}
J. Wang.
The scattering matrix for 0th order pseudodifferential
operators. Annales de l'Institut Fourier.
Tome 73, no 5, p.2185-2237, 2023.

\bibitem{Ya}
  D.R. Yafaev. The semiclassical limit of eigenfunctions of the Schr\"{o}dinger equation and the Bohr-sommerfeld quantization condition,
  revisited. Algebra i Analiz, 22(6):270–291, 2010.
   
\end{thebibliography}
%\addcontentsline{toc}{chapter}{Bibliographie}

\end{document}